\documentclass{vldb}
\usepackage{graphicx}
\usepackage{balance}
\usepackage{graphicx}
\graphicspath{ {./Figures/} }
\def\ps@empty{%
  \let\@mkboth\@gobbletwo\let\@oddhead\@empty\let\@oddfoot\@empty
  \let\@evenhead\@empty\let\@evenfoot\@empty}
\def\ps@plain{\let\@mkboth\@gobbletwo
     \let\@oddhead\@empty\def\@oddfoot{\reset@font\hfil\thepage
     \hfil}\let\@evenhead\@empty\let\@evenfoot\@oddfoot}

\usepackage{balance} 
\usepackage{graphicx}
\usepackage{times}
\usepackage{caption}

\usepackage{multirow}

\usepackage{hyperref}

\usepackage{amssymb}

\usepackage[normalem]{ulem}

\usepackage{subfigure}

\usepackage{times}
\usepackage{hyperref}

\usepackage{soul}

\usepackage[svgnames]{xcolor}

\usepackage[procnumbered,ruled, linesnumbered, vlined]{algorithm2e}

\usepackage{tcolorbox}

\SetAlFnt{\small}
\usepackage{xspace}
\newcommand{\nop}[1]{}

\newtheorem{lemma}{Lemma}

\newtheorem{definition}{Definition}
\newtheorem{example}{Example}

\usepackage{mathtools}

\usepackage{color}
\definecolor{Xiang}{rgb}{1,0,0}
\definecolor{Ahmed}{rgb}{0,0,1}

\usepackage[noabbrev]{cleveref}
\begin{document}
%
%
%
%

\vldbTitle{Topic-based Community Search over Spatial-Social Networks (Technical Report)}
\vldbAuthors{Ahmed Al-Baghdadi and Xiang Lian}
\vldbDOI{https://doi.org/10.14778/xxxxxxx.xxxxxxx}
\vldbVolume{13}
\vldbNumber{xxx}
\vldbYear{2020}


\title{Topic-based Community Search over Spatial-Social Networks (Technical Report)}

%
%

\markboth{Journal of \LaTeX\ Class Files,~Vol.~14, No.~8, August~2015}%
{Shell \MakeLowercase{\textit{et al.}}: Bare Demo of IEEEtran.cls for IEEE Journals}
%



\numberofauthors{1} 
\author{
\alignauthor
Ahmed Al-Baghdadi and Xiang Lian\\
       \textit{\large Department of Computer Science, Kent State University} \\
        Kent, OH 44242, USA \\
        \email {\{aalbaghd, xlian\}@kent.edu}}

\maketitle

\begin{abstract}
Recently, the community search problem has attracted significant attention, due to its wide spectrum of real-world applications such as event organization, friend recommendation, advertisement in e-commence, and so on. Given a query vertex, the community search problem finds dense subgraph that contains the query vertex. In social networks, users have multiple check-in locations, influence score, and profile information (keywords). Most previous studies that solve the CS problem over social networks usually neglect such information in a community.
In this paper, we propose a novel problem, named \textit{community search over spatial-social networks} (TCS-SSN), which retrieves community with high social influence, small traveling time, and covering certain keywords. In order to tackle the TCS-SSN problem over the spatial-social networks, we design effective pruning techniques to reduce the problem search space. We also propose an effective indexing mechanism, namely {\it social-spatial} index, to facilitate the community query, and develop an efficient query answering algorithm via index traversal. We verify the efficiency and effectiveness of our pruning techniques, indexing mechanism, and query processing algorithm through extensive experiments on real-world and synthetic data sets under various parameter settings.   

\end{abstract}


%
\section {Introduction}
With the increasing popularity of {\it location-based social networks} (e.g., Twitter, Foursquare, and Yelp), the community search problem has drawn much attention \cite{armenatzoglou2013general, zhang2013combining, fang2018spatial, yuan2016rsknn} due to its wide usage in many real applications such as event organization, friend recommendation, advertisement in e-commence, and so on. In order to enable accurate community retrieval, we need to consider not only social relationships among users on social networks, but also their spatial closeness on spatial road networks. Therefore, it is rather important and useful to effectively and efficiently conduct the community search over a so-called \textit{spatial-social network}, which is essentially a social-network graph integrated with spatial road networks, where social-network users are mapped to their check-in locations on road networks.

In reality, social-network users are very sensitive to post/propagate messages with different topics \cite{chen2015online}. Therefore, with different topics such as movie, food, sports, or skills, we may obtain different communities, which are of particular interests to different domain users (e.g., social scientists, sales managers, headhunting companies, etc.). In this paper, we will formalize and tackle a novel problem, namely \textit{topic-based community search over spatial-social networks} (TCS-SSN), which retrieves topic-aware communities, containing a query social-network user, with high social influences, social connectivity, and spatial/social closeness.

Below, we provide a motivation example of finding a group (community) of spatially/socially close people with certain skills (keywords) from spatial-social networks to perform a task together. \vspace{-1ex}

\setlength{\textfloatsep}{1pt}
\begin{figure}[t]
\centering\vspace{-3ex}
\includegraphics[width=10cm, height =5.5cm]{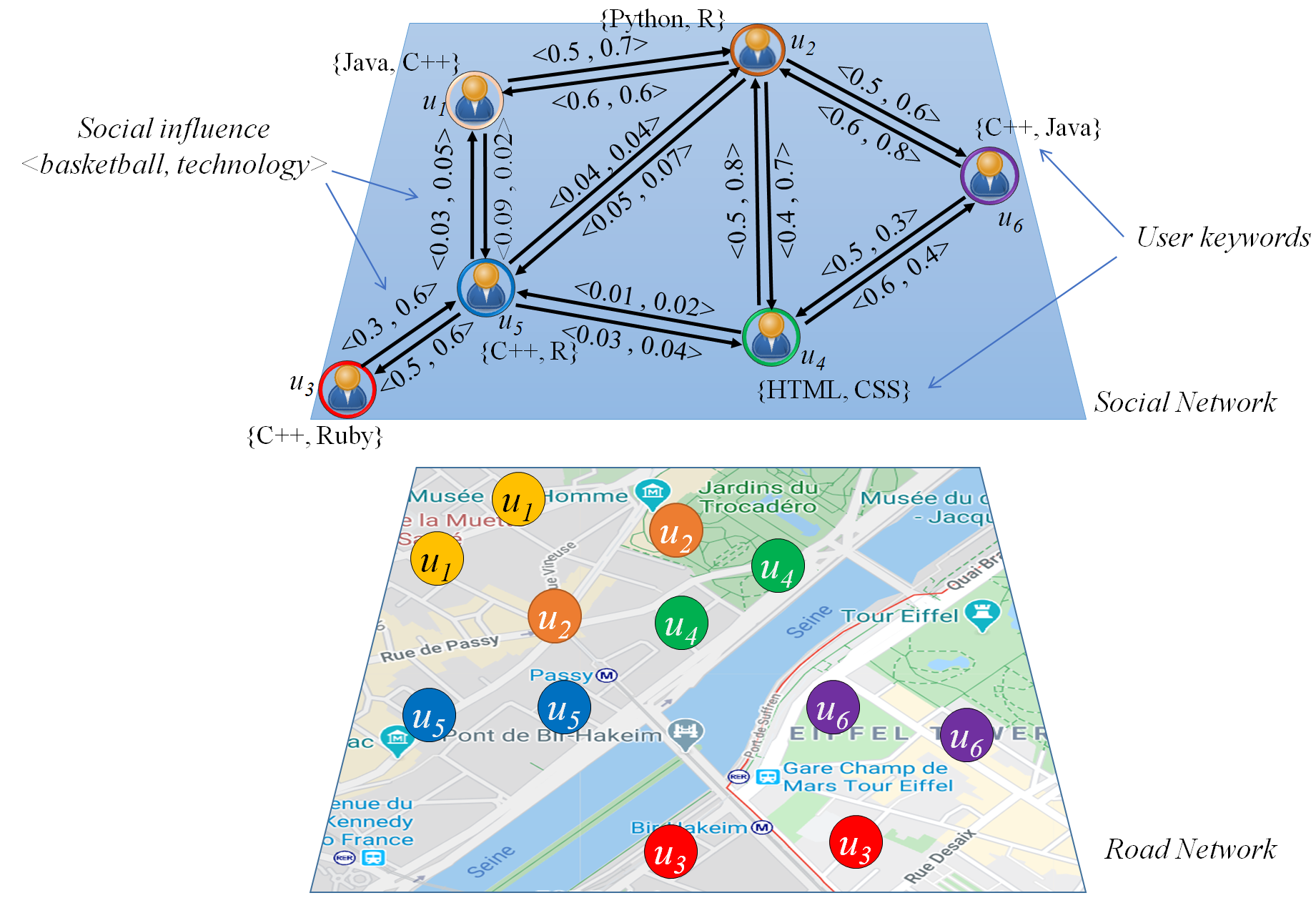}
\caption{An Example of Spatial-Social Networks.}
\label{fig:ssn}
\end{figure}

\begin{example} {\bf (Building a Project Team)} {\it}
Figure \ref{fig:ssn} illustrates an example of a \textit{spatial-social network}, $G_{rs}$, which combines social networks $G_s$ with spatial road networks $G_r$. In social networks $G_s$, users, $u_1\sim u_6$, are vertices, and edges (e.g., $e_{u_1,u_2}$) represent friend relationships between any two users. Each user (e.g., $u_1$) is associated with a set of keywords that represent his/her skills (e.g., programming language skills). Furthermore, each directed edge $e_{u_i,u_j}$ has a weight vector with respect to two topics, $(basketball,$ $technology)$, each weight representing the social influence of user $u_i$ towards user $u_j$ based on certain topic. For example, for the {\texttt{technology}} topic, the social influence of user $u_2$ on user $u_4$ is given by $0.7$, whereas the influence of user $u_4$ on user $u_2$ is $0.8$,
which shows asymmetric influence probabilities between users $u_2$ and $u_4$ on the \texttt{technology} topic.

Moreover, in road networks $G_r$, vertices are intersection points and edges indicate road segments containing connecting those intersection points. Each social-network user from social network $G_s$ has multiple check-in locations on the spatial network $G_r$.

In order to accomplish a programming project related to basketball websites, a project manager $u_2$ may want to find a voluntary (non-profit) team of developers who have the programming skills such as $\{Python, HTML, C++\}$ (i.e., topics), are socially and spatially close to each other, and highly influence each other on basketball topics. In this case, the manager can issue a TCS-SSN query over spatial-social networks $G_{rs}$ and call for a community of developers who can meet and complete the project task. 

In Figure \ref{fig:ssn}, although user $u_6$ has the query keyword $C++$ and high influence score with $u_2$, he/she resides in a place far away from $u_2$. Thus, $u_6$ will not be considered. Similarly, user $u_5$ will not be considered, since its influence score based on \texttt{basketball} and \texttt{technology} topic is very low. Therefore, in this running example, a community $\{u_2, u_1$, $u_4\}$ will be returned as potential team members.\qquad $\blacksquare$


\end{example}

As described in the example above, it is important that relationships among team members (programmers) to be high, so this will encourage them to join and better communicate with their friends. Also, we would like people with certain skills (topics) such as programming skills or front-end and back-end skills. Furthermore, we want programmers to reside within a certain road-network distance such that team members do not have to drive too long to meet.


Prior works on the community search usually consider the community semantics either by spatial distances only \cite{chen2015finding} and/or structural cohesiveness \cite{fang2018spatial}. They did not consider topic-aware social influences. Moreover, some works  \cite{huang2017attribute} considered communities based on topics of interests (e.g., attributes), however, topic-based social influences among users are ignored. 

In contrast, in this paper,  our TCS-SSN problem will consider the community semantics by taking into account degree of interests (sharing similar topics of interest) among users, degree of interactions (interacting with each other frequently), degrees of mutual influences (users who influence each other), structural cohesiveness (forming a strongly connected component on the social network), and spatial cohesiveness (living in places nearby on road networks). 

It is rather challenging to efficiently and effectively tackle the TCS-SSN problem, due to the large scale of spatial-social networks and complexities of retrieving communities under various constraints. Therefore, in this paper, we will propose effective pruning mechanisms that can safely filter out false alarms of candidate users and reduce the search space of the TCS-SSN problem. Moreover, we will design cost-model-based indexing techniques to enable our proposed pruning methods, and propose an efficient algorithm for TCS-SSN query answering via the index traversal.

Specifically, we make the following contributions in this paper.\vspace{-1ex}

\begin{enumerate}
\item We formalize the problem of the \textit{topic-based community search over spatial-social networks} (TCS-SSN) in Section \ref{sec:problem_def}.\vspace{-1.5ex}

\item We propose effective pruning strategies to reduce the search space of the TCS-SSN problem in Section \ref{sec:pruning_methods}.\vspace{-1.5ex}

\item We design effective, cost-model-based indexing mechanisms to facilitate the TCS-SSN query processing in Section \ref{sec:indexing}.\vspace{-1.5ex}

\item We propose an efficient query procedure to tackle the TCS-SSN problem in Section \ref{sec:Query_Answering}.\vspace{-1.5ex}

\item We demonstrate through extensive experiments the efficiency and effectiveness of our TCS-SSN query processing approach over real/synthetic data sets in Section
\ref{sec:ExperimentalEvaluation}.\vspace{-1ex}

\end{enumerate}

Section \ref{sec:related_work} reviews previous works on query
processing in social and/or road networks. Finally,
Section \ref{sec:conclusion} concludes this paper.


\section{Problem Definition}
\label{sec:problem_def}

\begin{table}[t!]
\centering{\scriptsize
\caption{\small Frequently used symbols and their descriptions.}
\label{table:symbols}
\begin{tabular}{||l|p{5.7cm}||}\hline
    {\bf Symbol} & {\bf Description} \\ \hline\hline
     $G_r, G_s,$ and $G_{rs}$ & a spatial network, a social network, and a spatial-social network, respectively\\ \hline
     $S$ & a set (community) of social-network users\\ \hline
     $u$, $v$, $u_j$, or $v_j$ & a social-network user \\ \hline
     $u.key$ & a vector of possible keywords associated with user $u_j$\\ \hline
     $u.L$ & a set of check-in locations, $u.loc_i$, by user $u$\\\hline
     $e_{u, v}$ & a directed edge from user $u$ to user $v$\\ \hline
     $e_{u, v}.\mathcal{T}$ & a vector of topics of interests associated with edge $e_{u, v}$ \\\hline
     $sup(e)$ & the support of an edge $e$\\\hline
     $tp^{j}_{u, v}$ & a weight probability on the topic $j$ from user $u$ to $v$\\ \hline
     $infScore(.,.)$ & an influence score function\\\hline
     $dist_{RN}(u.loc_i, v.loc_j)$ & the shortest road-network distance between 2 locations  \\ \hline
     $avg\_dist_{RN}(u, v)$ & the average road-network distance between users $u$ and $v$ \\ \hline
     $dist_{SN}(u, v)$ & No. of hops between users $u$ and $v$ on social networks\\ \hline
      $\mathcal{P}_{RN}$ & a set of $l$ road-network pivots, $rpiv_i$\\\hline
      $\mathcal{P}_{SN}$ & a set of $h$ social-network pivots, $spiv_i$\\\hline
      $\mathcal{P}_{index}$ & a set of $\iota$ index pivots\\\hline
\end{tabular}
}
\end{table}

In this section, we provide formal definitions and data models
for social networks, spatial networks, and their combination, spatial-social networks, and then define our novel query of {\bf t}opic-based {\bf c}ommunity {\bf s}earch over {\bf s}patial-{\bf s}ocial {\bf n}etworks (TCS-SSN).

\subsection{Social Networks}
We formally define the data model for social networks, as well as structural cohesiveness and social influence in social networks.

\begin{definition}
{\bf (Social Networks, $G_s$)} A social network, $G_s$, is a triple
$(V(G_s),E(G_s),\phi(G_s))$, where $V(G_s)$ is a set of $M$ users
$u_{1}$, $u_{2}$, $\dots$, and $u_M$, $E(G_s)$ is a set of edges
$e_{u,v}$ (friendship between two users $u$ and $v$), and
$\phi(G_s)$ is a mapping function: $V(G_s) \times V(G_s)
\rightarrow E(G_s)$.\label{def:social_networks}
\end{definition}

In Definition \ref{def:social_networks}, each user, $u_j$ (for $1\leq j\leq M$), in the social network $G_s$ is associated with a vector of possible keywords (or skills) $u_j.key = (key^{j}_1, key^{j}_2, \dots, key^{j}_{|key|})$.

Each edge (friendship), $e_{u,v}$ ($\in E(G_s)$), in social networks is associated with a vector of topics of interest $e_{u,v}.\mathcal{T} = (tp^1_{u,v}, tp_{u,v}^2,$ $\dots, tp^{|\mathcal{T}|}_{u,v})$,
where $tp_{u,v}^j$ is the influence probability (weight) of interested topic $j$ and $|\mathcal{T}|$ is the size of the topic set $e_{u,v}.\mathcal{T}$.


\vspace{0.5ex}\noindent \textbf{Modeling Structural Cohesiveness}: Previous works usually defined the community as a subgraph in social networks $G_s$ with high structural cohesiveness. In this paper, to capture structural cohesiveness in $G_s$, we consider the connected $(k, d)$-truss \cite{huang2017attribute}. 

Specifically, we first define a \textit{triangle} in $G_s$, which is a cycle of length 3 denoted as $\triangle_{u_a u_b u_c}$, for user vertices $u_a, u_b, u_c \in V(G_s)$; the support, $sup(e)$, of an edge $e \in E(G_s)$ is given by the number of triangles containing $e$ in $G_s$ \cite{wang2012truss}. Then, the connected $(k, d)$-truss is defined as follows.

\begin{definition}(\textbf{Connected $(k, d)$-Truss} \cite{huang2017attribute}):
\textit{
Given a graph $G_s$, and an integer $k$, a connected subgraph $S \in G_s$ is called a $(k, d)$-truss if two conditions hold: (1) $\forall e\in E(S), sup(e)\geq (k-2)$, and (2) $\forall u, v \in V(S), dist_{SN}(u, v)$ $< d$, where $sup(e)$ is the number of triangles containing $e$ and $dist_{SN}(u, v)$ is the shortest path distance (the minimum number of hops) between users $u$ and $v$ on social networks.
}
\label{def:k-d-truss}
\end{definition}


\vspace{0.5ex}\noindent \textbf{Modeling Social Influences}: Now, we discuss the data model for social influences in social networks. Each edge $e_{u,v}$ is associated with a vector, $\mathcal{T} = (tp^1_{u,v}, tp_{u,v}^2,$ $\dots, tp^{|\mathcal{T}|}_{u,v})$, of influence probabilities on different topics. Further, we denote the path $path_{u,v}$ as a path on the social network connecting two users $u$ and $v$ such that $path_{u,v}= ( u = a_1\to a_2\to \dots \to a_{|path_{u, v}|} = v)$.
It is worth noting that, Barbieri et al. \cite{barbieri2013topic} extended the classic IC and LT models to be topic-aware and introduced a novel topic-aware influence-driven propagation model that is more accurate in describing real-world cascades than standard propagation models.
In fact, users have different interests and items have different characteristics, thus, we follow the text-based topic discovery algorithm \cite{barbieri2013topic, chen2015online} to extract user's interest topics and their distribution on each edge.
Specifically, for each edge $e_{u, v}$, we obtain an influence score vector, for example, (basketball:0.1, technology:0.8), indicating that the influence probabilities of user $v$ influenced by user $u$ on topics, basketball and technology, are 0.1 and 0.8, respectively.

Below, we define the influence score function.

\begin{definition}(\textbf{Influence Score Function} \cite{chen2015online}).
Given a social-network graph $G_s$, a topic vector $\mathcal{T}$, and two social-network users $u, v \in V(G_s)$, we define the influence score from $u$ to $v$ as follows:
\begin{equation}
\hspace{-2ex}infScore(u, v|\mathcal{T}) = \max_{ \forall path_{u, v} \in G_s}\{ infScore(path_{u, v}|\mathcal{T})\}.
\end{equation}

For two vertices $u, v\in V(G_s)$ such that $path_{u,v}= ( u = a_1 \to a_2\to \dots \to a_{|path_{u, v}|} = v)$, and a topic vector $\mathcal{T}$,
we define the influence score on $path_{u, v}$ as follows:\vspace{-2ex}

\begin{eqnarray}
 infScore(path_{u, v}| \mathcal{T})= \prod_{i=1}^{|path_{u, v}|-1}{f(a_i, a_{i+1}| \mathcal{T})},
 \label{eq:infScore_TwoVertices}
\end{eqnarray}
where $f(a_i, a_{i+1}| \mathcal{T})$ is the influence score from $a_i$ to $a_{i+1}$ of the two adjacent vertices $a_i$ and $a_{i+1}$ based on the query topic vector $\mathcal{T}$. We compute the influence score between any two adjacent vertices $a_i$ and $a_{i+1}$ as follows:\vspace{-2ex}

\begin{equation}
f(a_i, a_{i+1}|\mathcal{T}) = \sum_{j= 1}^{|\mathcal{T}|}{tp_{u, v}^j \cdot \mathcal{T}^j},\vspace{-1ex}
\end{equation}

\noindent where $tp_{u,v}^j$ is the weight probability on topic $j$, $\mathcal{T}^j$ is the $j$-th query topic, and $|\mathcal{T}|$ is the length of the topic set $\mathcal{T}$.
\label{def:influnceScoreFunction}
\end{definition}

In Definition \ref{def:influnceScoreFunction}, we define the influence score between any two users in the social network $G_s$. Given a topic vector $\mathcal{T}$ and a subset $S \subseteq G_s$, we define the influence score between subgraph $S$ and a user $v$ as follows:
\begin{equation}
    infScore(S, v|\mathcal{T}) = \min_{\forall u\in S }\{ infScore(u, v|\mathcal{T})\},
\end{equation}
where $infScore(u, v|\mathcal{T})$ is defined in Eq. (\ref{eq:infScore_TwoVertices}).

Note that, the pairwise influence (or mutual influence) in our TCS-SSN problem indicates the influence of one user on another user in the community. In particular, the pairwise influence is not symmetric, in other words, for two users $u$ and $v$, the influence of $u$ on $v$ can be different from that of $v$ on $u$. Thus, in our TCS-SSN community definition, we require both influences, from $u$ to $v$ and from $v$ to $u$, be greater than the threshold $\theta$ (i.e., mutual influences between $u$ and $v$ are high), which ensures high connectivity or interaction among users in the community. 
   Other metrics such as pairwise keyword similarity \cite{bhattacharyya2011analysis} (e.g., Jaccard similarity) are usually symmetric (providing a single similarity measure between two users), which cannot capture mutual interaction or influences. Most importantly, users $u$ and $v$ may have common keywords/topics, however, it is possible that they may not have high influences to each other in reality.

\subsection{Spatial Road Networks} 
 
Next, we give the formal definition of spatial road networks. 

\begin{definition}
{\bf(Spatial Road Networks, $G_r$)} A spatial road network, $G_r$, is represented by a
triple $(V(G_r),E(G_r),\phi(G_r))$, where $V(G_r)$ is a set of $N$
vertices $w_{1}$, $w_{2}$, $\dots$, and $w_N$, $E(G_r)$ is a set of
edges $e_{j,k}$ (i.e., roads between vertices $w_j$ and $w_k$), and
$\phi(G_r)$ is a mapping function: $V(G_r) \times V(G_r)
\rightarrow E(G_r)$.\label{def:spatial_networks}
\end{definition}
In Definition \ref{def:spatial_networks}, road network $G_r$ is modeled by a graph, with edges as roads and vertices as intersection points of roads.

\subsection{Spatial-Social Networks}
In this subsection, we define spatial-social networks, as well as the spatial cohesiveness over spatial-social networks.

\begin{definition}
{\bf (Spatial-Social Networks, $G_{rs}$)} A spatial-social network, $G_{rs}$,
is given by a combination of spatial road networks $G_r$ and social
networks $G_s$, where users
$u_j$ on social networks $G_s$ are located on some edges of spatial road
networks $G_r$.
\label{def:spatial_social_networks}
\end{definition}

From Definition \ref{def:social_networks}, each social-network user, $u_j \in G_s$ (for $1\leq j\leq M$), is associated with a 2D location on the spatial network $u_j.L$, where $u_j.L = \{ u^j_{(loc_1, time_1)}, \dots, (u^j_{(loc_{|u.L|}, time_{|u.L|})}\}$, where $u^j_{loc_i}$ has its spatial coordinates $(x^j_i, y^j_i)$ along $x-$ and $y-$axes, respectively on $G_r$ at timestamp $u^j_{time_i}$.


\vspace{0.5ex}\noindent \textbf{Modeling Spatial Cohesiveness}: Next, we discuss modeling of spatial cohesiveness over spatial-social networks.
In real-world social networks, users change their locations frequently due to mobility. As a result, users' spatially close communities change frequently as well \cite{fang2018spatial}.
Social-network users' check-in information can be recorded with the help of GPS and WiFi technologies.
To measure the spatial cohesiveness, we define an average spatial distance function, $avg\_dist_{RN}(.)$. The average spatial distance function utilizes social-network users' locations $u.loc$ on the spatial network to measure the spatial cohesiveness.

\begin{definition}(\textbf{The Average Spatial Distance Function}).
Since each social-network user $u \in V(G_s)$ has multiple locations on spatial networks $G_r$, $u.loc$, at different timestamp $time$, we define the shortest path distance between any two users $u, v\in V(G_s)$ on the spatial networks as follows:
\begin{eqnarray}
\label{eq:spatialShortestPath}
    avg\_dist_{RN}(u, v) = \frac{\sum _{\forall u.loc_i}{\sum_{\forall v.loc_j}{dist_{RN}(u.loc_i, v.loc_j)}}}{|u.L| \cdot |v.L|},
    \end{eqnarray}
where $|u.L|$ is the number of check-ins by user $u$, and $dist_{RN}(.,.)$ is the shortest path distance between two road-network locations.
\label{def:maxSpatialDistance}
\end{definition}

\nop{
Furthermore, we define the average road-network distance between a set $S$ of social-network users and a user $v$ as follows:

\begin{equation}
    dist_{RN}(S,v) = \min_{\forall u \in S}\{avg\_dist_{RN}(u, v)\},
    \label{eq:SpatialDistance_S_to_v}
\end{equation}
where $dist(S,v)$ is the average shortest path distance between a set of social-network users $S$ and a user $v$, and $avg\_dist_{RN}$ is defined in Eq.~\eqref{eq:spatialShortestPath}
}


\subsection{\underline{T}opic-based \underline{C}ommunity \underline{S}earch over \underline{S}patial-\underline{S}ocial \underline{N}etwork (TCS-SSN)}

In this subsection, we first propose a novel spatial-social structure, ss-truss, and then formally define our TCS-SSN problem.

\vspace{0.5ex}\noindent {\bf Spatial-Social Structure, ss-truss.} In this work, we consider both spatial and social networks to produce compact communities with respect to spatial cohesiveness, social influence, structural cohesiveness, and user keywords.
We propose a novel spatial-social $(k, d, \sigma, \theta)$-truss, or ss-truss.

\begin{definition}({\bf Spatial-Social $(k, d, \sigma, \theta)$-Truss}, ss-truss).
\textit{
Given a spatial-social network $G_{rs}$, a query topic set $\mathcal{T}_q$, integers $k$ and $d$, a spatial distance threshold $\sigma$, and an influence score threshold $\theta$, we define the spatial-social $(k, d, \sigma, \theta)$-truss, or ss-truss, as a set, $S$, of users from the social network $G_s$ such that:\vspace{-1ex}
\begin{itemize}
    \item $S$ is a $(k, d)$-truss (as given in Definition \ref{def:k-d-truss});\vspace{-1ex}
    \item the average spatial distance between any two users $u$ and $v$ in $S$ is less than $\sigma$, $u, v \in S| avg\_dist_{RN}(u, v) < \sigma$, and;\vspace{-1ex}
    \item the influence score $infScore(.|.)$, $\{ \forall u, v \in S: \exists path_{u, v} \in S, infScore(path_{u, v}|\mathcal{T}_q) \geq \theta$.\vspace{-1ex}
\end{itemize}
}\label{def:ss-truss}
\end{definition}

Note that, the ss-truss satisfies the nested property that: if $k'\leq k$, $d’ \geq d$, $\sigma' \leq \sigma$, and $\theta’\leq \theta$ hold, then we have: ($k$, $d$, $\sigma$, $\theta$)-truss is a subgraph of some ($k’$, $d’$, $\sigma’$, $\theta’$)-truss.

Now, we define our novel query topic-based community search over spatial-social networks.

\begin{definition}({\bf Topic-based Community Search Spatial-Social Community, TCS-SSN}).
Given a spatial-social network $G_{rs}$, a query user $q$, a keyword set query $K_q$, and a topic query set $\mathcal{T}_q$,
the topic-based community search over spatial-social networks (TCS-SSN) retrieves a maximal set, $S$, of social-network users such that:\vspace{-1ex}
\begin{itemize}
    \item $q \in S$;\vspace{-1ex}
    \item $S$ is a $(k, d, \sigma, \theta)$-truss, and;\vspace{-1ex}
    \item $\forall u \in S, u.key \cap K_q \neq \emptyset$.\vspace{-1ex}
\end{itemize}
\label{def:TCS-SSN}
\end{definition}

\nop{
\begin{definition}{({\bf Temporal Spatial-Social Events})}.
We define temporal spatial social events over the temporal spatial-social networks as follows:
\begin{itemize}
    \item \textbf{social event}: a social network user triggers a social event, if the user substantially changed his/her social influence or its network structure.
    \item \textbf{spatial event}: a social network user triggers a spatial event if the user significantly changed his/her spatial location.
    \item \textbf{temporal event}: a social network user triggers a temporal event if the user joined the social community $S_A$ in $GsT1$ and changed his social community to be in $S_B$ in $G_sT2$.
\end{itemize}
\end{definition}
}

\noindent{\bf Discussions on the Parameter Settings:}
Note that, parameter $\theta$ ($\in$[0, 1]) is an influence score threshold that specifies the minimum score that any two users influence each other based on certain topics in the user group $S$. Larger $\theta$ will lead to a user group $S$ with higher social influence.

The topic query set, $\mathcal{T}_q$, contains a set of topics specified by the user. The influence score between any two users in the user group $S$ is measured based on topics in $\mathcal{T}_q$. The larger the topic set query $\mathcal{T}_q$, the higher the influence score among users in the resulting community $S$.

The parameter $\sigma$ controls the maximum (average) road-network distance between any two users in the user group $S$, that is, any two users in $S$ should have road-network distance less than or equal to $\sigma$. The larger the value of $\sigma$, the farther the driving distance between any two users in the community community $S$.

The parameter $d$ limits the maximum number of hops between any two users in the user group $S$ on social networks. The larger the value of $d$, the larger the diameter (or size) of the community $S$. 
   
 The integer $k$ controls the structural cohesiveness of the community (subgraph) $S$ in social networks. That is, $k$ is used in $(k, d)$-truss to return a community $S$ with each connection (edge) ($u$, $v$) endorsed by $(k - 2)$ common neighbors of $u$ and $v$. The larger the value of $k$, the higher the social cohesiveness of the resulting community $S$.


The keyword query set $K_q$, is a user-specified parameter, which contains the keywords or skills a user $u$ must have in order to be included in the community. In real applications (e.g., Example 1, each user in the resulting community $S$ must have at least one keyword in $K_q$.

To assist the query user with setting the TCS-SSN parameters, we provide the guidance or possible fillings of parameters $\theta$, $\mathcal{T}_q$, $\sigma$, $d$, and $k$, such that the TCS-SSN query returns a non-empty answer set. Specifically, for the influence threshold $\theta$, we can assist the query user by providing a distribution of influence scores for pairwise users, or suggesting the average (or x-quantile) influence score of those user groups selected in the query log. 
To suggest the topic query set $\mathcal{T}_q$, we can give the user a list of topics from the data set, and the user can choose one or multiple query topics of one's interest. 
Furthermore, to decide the road-network distance threshold $\sigma$, we can also show the query user a distribution of the average road-network distance between any neighbor users (or close friends) on social networks.
In addition, we suggest the setting of value $k$, by providing a distribution of supports, $sup(e)$, on edges $e$ (between pairwise users) of social networks, and let the user tune the social-network distance threshold $d$, based on the potential size of the resulting subgraph (community). Finally, we assist the query user setting the keyword query set $K_q$ by providing a list of frequent keywords appearing in profiles of users surrounding the query issuer $q$.

\noindent {\bf Challenges:} The straightforward approach to tackle the TCS-SSN problem is to enumerate all possible social-network users, check  query predicates on spatial-social networks (as given in Definition \ref{def:TCS-SSN}), and return TCS-SSN query answers. However, this method incurs high time complexity, since the number of possible users in a community is rather large. 
Although some of users with unwanted keywords can be directly discarded, still there will be a very large group of users satisfying the query keyword set. Thus, in the worst case, there is an exponential number of possible combination of users groups. For each user group, spatial-cohesiveness, structural-cohesiveness, and influence score have to be measured to obtain final TCS-SSN answers, which is not efficient. Applying such measures to many group of users may not be even feasible with nowadays social networks containing millions of nodes and edges.

Therefore, in this work, we will design effective pruning strategies to reduce the search space of the TCS-SSN problem. Then, we will devise indexing mechanisms and develop efficient TCS-SSN query answering algorithms by traversing the index. 

\section{Pruning Methods}
\label{sec:pruning_methods}
In this section, we propose effective pruning techniques that utilize the topic-based community search properties to reduce the search space and facilitate the online community search query processing.

\subsection{Spatial Distance-Based Pruning}
\label{subseq:SpatialDistance-BasedPruning}
For any ss-truss community $S$, the average spatial distance between any pair of users is less than $\sigma$ (as given in Definition \ref{def:ss-truss}).
Based on that, for any two social-network user, if the average spatial distance between their check-in locations in the spatial network is greater than $\sigma$, then they cannot be in the same community.
We propose our spatial distance-based pruning that prunes false alarms w.r.t. $\sigma$ threshold in the ss-truss.

\nop{
\begin{lemma}(\textbf{Spatial Distance-Based Pruning}).
Given a spatial-social network $G_{rs}$, a candidate spatial-social community (ss-truss) $S$ containing the query vertex $q$, and a candidate vertex $v \in V(G_s)$ to be in $S$, the vertex $v$ can be safely pruned if $\exists u \in S | avg\_dist_{RN}(u, v) \geq \sigma$, where $avg\_dist_{RN}(.,.)$ is defined in Eq.~\eqref{eq:spatialShortestPath}.
\label{lem:SpatialDistance-BasedPruning}
\end{lemma}
}
Intuitively, if the average spatial distance between a vertex $v$ and a candidate vertex $u$ is greater than $\sigma$, it means that user $v$ resides in a place far from $v$. By the lemma, $v$ can be discarded.
However, the computation of the average spatial distance is costly. Next, we present our method of computing the average spatial distance between social-network users.

\noindent \textbf{Computing the Average Spatial Distance}:
For two social-network vertices $u$ and $v$, the average spatial road distance is computed by applying Eq.~\eqref{eq:spatialShortestPath}.
Since each social-network user may have multiple check-in locations on the spatial network, Eq.\eqref{eq:spatialShortestPath} enumerates all possible shortest path combinations between the check-in locations of the two users.

From Figure \ref{fig:ssn}, assume that we would like to compute the average spatial shortest path distance between $u_6$ and $u_4$.
Since each user has 2 check-in locations, 4 shortest path distance computations on road networks are required.
Clearly, Eq.~\eqref{eq:spatialShortestPath} cannot be applied to large graphs due to its high time complexity.
Thus, we will develop a pruning method to reduce the computation cost and tolerate real-world large graphs.

To reduce computational costs, we avoid the computation of the exact average spatial distance between two users by estimating the lower bound of the average spatial distance between them.
\begin{lemma}
For any user $u$ in the ss-truss community $S$, and a user $v$ to be in $S$, if the lower bound $lb\_avg\_dist_{RN}(u, v)$ of the average shortest path distance is greater than the spatial distance threshold, $lb\_avg\_dist_{RN}(u,v) > \sigma$, then user $v$ cannot be in $S$ and can be safely pruned.
\label{lem:UpperBoundoftheAverageSpatialDistance}
\end{lemma}

We will utilize the triangle inequality \cite{al2019efficient} to estimate the lower bound $lb\_avg\_dist_{RN}(.,.)$ of the average spatial distance between any two vertices. We rely on the spatial distance offline pre-computation of road network pivots $\mathcal{P}_{R}$ to estimate the lower bound of the average spatial distance between any two social-network users. We offline pre-compute the shortest path distance from each user's check-in locations $u.loc_i( 1 \leq i \leq |u.L|)$ to all pivot locations $\mathcal{P}_{RN}= \{ rpiv_1, \dots, rpiv_{l}\}$. 

By the triangle inequality, we have: $dist_{RN}(u.loc_i, v.loc_j) \geq |dist_{RN}(u.loc_i, piv_k) - dist_{RN}(piv_k, v.loc_j)|$, where $dist_{RN}(u.loc_i,$ $piv_k)$ (or $dist_{RN}(piv_k, v.loc_j)$) is the shortest path distance on the road network between the $i$-th location of user $u$ (or the $j$-th location of user $v$) and the $k$-th pivot, $ 1 \leq i \leq |u.L|$, $ 1 \leq j \leq |v.L|$, and $ 1 \leq k \leq l$. Then, at the query time, we utilize this triangle inequality property to estimate the average spatial distance lower bound, $lb\_avg\_dist_{RN}(u, v)$, of any two social-network users $u$ and $v$ in Eq. (\ref{eq:spatialShortestPath}).

{\scriptsize
\begin{eqnarray}
  \hspace{-6ex}  
  && avg\_dist_{RN}(u, v)\label{eq:ub_d_rsp}\\ \nonumber
  \hspace{-6ex}  & \geq & \max_{k=1}^{l}\left\{
     \frac{\sum_{i=1}^{|u.L|} \sum_{j=1}^{|v.L|} |(dist_{RN}(u.loc_i, rpiv_k) - dist_{RN}(rpiv_k, v.loc_j))| }{|u.L|. |v.L|}
    \right\}\\\nonumber
  \hspace{-6ex}  &=& lb\_avg\_dist_{RN}(u, v).
\end{eqnarray}
}

\noindent where $dist_{RN}(\cdot, \cdot)$ is the shortest path distance on the road network, $l$ is the number of road-network pivots $\mathcal{P}_{RN}$, and $|u.L|$ (or $|v.L|$) is the number of check-in locations by user $u$ (or $v$).

\subsection{Influence Score Pruning} 
\label{subsec:InfluenceScorePruning}
In Definition \ref{def:ss-truss}, for a set $S$ of social-network users to be an ss-truss, the influence score between any pair of users should be greater than a certain threshold $\theta$. This ss-truss property ensures that the resulting communities have high influence scores, that is, users in communities highly influence each other. In the sequel, we propose a pruning method that utilizes this property to reduce the search space by filtering out users with low influence score.


    For a user $v$ to be in a spatial-social community (ss-truss) $S$, based on influence score, the influence score between $v$ and each vertex in $S$ has to be greater than or equal to $\theta$.

    We propose an effective influence score pruning with respect to influence score upper bounds below.

\begin{lemma}(\textbf{Influence Score Pruning}).
\textit{
Given a social network $G_s$, a spatial-social community (ss-truss) $S$, a topic query $\mathcal{T}_q$, and a candidate vertex $v$ to be in $S$, the vertex $v$ can be safely pruned if there exists a vertex $u\in S$ such that $ub\_infScore(u, v| \mathcal{T}_q)< \theta$.
\label{lem:UB_InfluenceScorePruning}
}
\end{lemma}

For each user $u$ in the social network $G_s$, we utilize the influence score upper bounds to efficiently prune false alarms. Next, we describe our method of computing a tight upper bound of the influence score between any two vertices. 

\noindent {\bf The Computation of the Influence Score Upper Bound}:
We denote the in-degree of a vertex $u$ as $u.deg_{in}$, where $u.deg_{in}$ is a set of users $v\in V(G_s)$ such that $e_{v,u}\in E(G_s)$, and $u.deg_{out}$ is the out-degree as a set of users $v\in V(G_s)$ such that $e_{u,v}\in E(G_s)$.
We denote $ub\_inf^{in}(u)$ and $ub\_inf^{out}(u)$ as the upper bound of in/out-influence of the vertex $u$.
We compute the $ub\_inf^{in}(.)$ and $ub\_inf^{out}(.)$ as follows:
\begin{eqnarray}
   &&ub\_inf^{in}(u|\mathcal{T})\\ \nonumber 
   &=& \forall _{v\in u.deg_{in}} \forall_{t \in \mathcal{T}} \{ \max\{tp^t_{v,u}\}, \dots, \max\{tp^{|\mathcal{T}|}_{v,u}\}\}
   \label{eq:userIN_influenceScoreUperBound}
\end{eqnarray}
\begin{eqnarray}
   && ub\_inf^{out}(u|\mathcal{T}) \\ \nonumber
   &=& \forall _{v\in u.deg_{out}} \forall_{t \in \mathcal{T}} \{ \max\{tp^t_{u, v}\}, \dots, \max\{tp^{|\mathcal{T}|}_{u, v}\}\}
   \label{eq:userOut_influenceScoreUperBound}
\end{eqnarray}

For any two nonadjacent vertices $u, v \in V(G_s)(i.e., e_{u,v} \notin E(G_s))$, we estimate the upper bound of the influence score from $u$ to $v$ as follows:
\begin{eqnarray}
   ub\_infScore(u, v| \mathcal{T})
   = ub\_inf^{out}(u|\mathcal{T} ) \cdot ub\_inf^{in}(v|\mathcal{T}).
   \label{eq:ub_infScore}
\end{eqnarray}

Estimating the upper bound of the influence score is very critical for the influence score pruning to perform well. In Eq.~\eqref{eq:ub_infScore}, we utilize one hop friends to estimate the upper bound of the influence score. This method has proven to be effective and we will show in the experimental evaluation, Section \ref{sec:ExperimentalEvaluation}.

\subsection{Structural Cohesiveness Pruning}
\label{subsec:StructuralCohesivenessPruning}
The ss-truss communities have high structural cohesiveness. From Definition \ref{def:k-d-truss}, the support of an edge in ss-truss community $S$ has to be greater than or equal $k-2$, $sup(e) \geq k-2$.
We refer $\Phi(u), u\in V(G_s)$, as the maximum support of an edge induced by $u$, mathematically,

\begin{eqnarray}
\label{eq:phi}
 \Phi(u) = \max_{\forall v \in u.deg} \begin{dcases} 
    sup(u,v) & if (v \in u.deg_{out}); \\
     sup(v, u) & if (v \in u.deg_{in}),
\end{dcases}
\end{eqnarray}
where $ u.deg = \{u.deg_{in} \cup u.deg_{out}\}$.
\begin{lemma} ({\bf Structural Cohesiveness Pruning}).
Given a social network $G_s$, a spatial-social community (ss-truss) $S$, and a candidate vertex $v$ to be in $S$, vertex $v$ can be directly pruned, if $\Phi(v) < k-2 $.
\label{lem:StructuralCohesivenessPruning}
\end{lemma}

\nop{
\begin{proof}
If $\Phi(v) < k-2 $, that means the maximum support of any edge induced by $v$ is less $k - 2$. From Definition \ref{def:TCS-SSN} the vertex $v$ cannot be in $S$. Based on that, $v$ can be safely discarded.
\end{proof}
}

Computing the edge support is a key issue to apply Lemma \ref{lem:StructuralCohesivenessPruning}. In this regard, we rely on Wang et al. \cite{wang2012truss} to compute the maximum edge support for all edges in the graph, $sup(e), \forall e \in E(G_s)$, in $O(E(G_s)^{1.5})$.

\sloppy
\subsection{Social Distance-Based Pruning}
\label{subseq:SocialDistance-BasedPruning}
For a spatial-social community $S\in G_s$, Definition \ref{def:ss-truss} ensures that
for any two vertices $u, v\in S$, the shortest path distance connecting $u$ and $v$ over the social network $G_s$ must be less than $d$, $dist_{SN}(u, v) < d, \forall u,v \in S$.
In the social distance-based pruning, we filter out vertices with distances greater than $d$ from the candidate set $S$.
\begin{lemma} (\textbf{Social Distance-Based Pruning}).
Given a social network $G_s$, a spatial-social community (ss-truss) $S$, and a candidate vertex $v$ to be in $S$, the vertex $v$ can be directly filtered out if $lb\_dist_{SN}(S, v)\geq d$, where $lb\_dist_{SN}(S, v) = \min_{\forall u\in S}\{ lb\_dist_{SN}(u, v)\}$.
\label{lem:SocialDistance-BasedPruning}
\end{lemma}
 
\noindent {\bf The Computation of the Social Distance Lower Bound:}
 For a two social-network users $u$ and $v$, the social network distance is the minimum number of hops connecting $u$ and $v$. The lower bound of the social-network distance between $u$ and $v$ can be computed by utilizing triangle inequality. We offline pre-compute the social-network distance from user to all social-network pivots $\mathcal{P}_{SN}= \{ spiv_1, \dots ,spiv_h\}$. At query time, use triangle inequality to estimate the social-network distance between any two social-network users $u$ and $v$ as follows:
  \begin{eqnarray}
   dist_{SN}(u, v) \nonumber
  &\geq& \max_{k=1}^h \{   dist_{SN}(u, spiv_k) - dist_{SN}(v, spiv_k) \}\\ \noindent
  & = & lb\_dist_{SN}(u, v),
 \end{eqnarray}
 where $dist_{SN}(u,spiv_k)$ is the shortest path social-network distance between user $u$ and the $k$-th social-network pivot, $(1 \leq k \leq h)$, and $h$ is the number of the social-network pivots $\mathcal{P}_{SN}$.
 
 \nop{
\begin{proof}
If $ub\_dist_{SN}(S, v) \geq d$, then there is at least one vertex $u\in V(S)$, such that $ub_dist_{SN}(u, v) \geq d$, that indicates $dist_{SN}(u, v) \geq d$. By Definition \ref{def:ss-truss}, the vertex $v$ violates the maximum social distance property and cannot join $S$, thus it can be directly pruned.
\end{proof}
}

\subsection{Keyword-based Pruning}
For a user $v$ to join the candidate ss-truss community $S$, the user keyword set $v.key$ has to cover at least one keyword in the keyword query set $K_q$. If the candidate vertex $v.key$ shares no keyword with the query set $K_q$, then $v$ can be discarded.

\begin{lemma}(\textbf{Keyword-based Pruning}).
Given a social-network graph $G_s$,
a spatial-social network set $S$,
a keyword query set $K_q$,
and a user $v$ to be in $S$, user $v$ can be safely pruned, if $v.key \cap K_q= \emptyset$.
\label{lem:Keyword-basedPruning}
\end{lemma}

\nop{
\begin{proof}
If $v.key \cap K_q= \emptyset$, there exists no keyword $k \in v.key$ such that $k\in K_q$. By Definition \ref{def:SocialNetworkKeywordConstrain}, it is safe to reject $v$.
\end{proof}
}

\section{Indexing Mechanism}
\label{sec:indexing}
\subsection{Social-Spatial Index, $\mathcal{I}$, Structure}
\label{subsec:Social-SpatialIndex}
We build our social-spatial index $\mathcal{I}$ over social-network vertices $V(G_s)$. Specifically, we utilize information from both spatial and social networks to partition the social network vertices into subgraphs. The subgraphs can be treated as leaf nodes of the index $\mathcal{I}$. Then, connected subgraphs in leaf nodes are recursively grouped into non-leaf nodes, until a final root is obtained.

\noindent{\bf Leaf Nodes.}
Each leaf node in the social-spatial index $\mathcal{I}$ contains social-network users $u$. 
Each user $u$ in leaf nodes is associated with a vector of the user's 2D check-in locations $u.L$, a set of keywords $u.key$, a vector of the maximum out-influence topics $u.inf^{out}$, a vector of the maximum in-influenced topics $u.inf^{in}$, and the minimum value of edge support associated with the user $u$, $\Phi(u)$.
To save the space cost, we hash each keyword $k \in u.key$ into a position in a bit vector $u.V_{key}$.

Furthermore, we choose $h$ social-network pivots $\mathcal{P}_{SN}= \{spiv_1, spiv_2, \dots$, and $spiv_h$ \} in $G_s$.
Similarly, we choose $l$ road-network pivots $\mathcal{P}_{RN}= \{rpiv_1, rpiv_2, \dots$, and $rpiv_l$\} in $G_r$.
Each social-network user $u$ in leaf nodes maintains its social-network distance to the social-network pivots, that is, $dist_{SN}(u, spiv_i)(1\leq i\leq h)$. 
The case of road-network pivots $avg\_dist_{RN}(u, rpiv_j)(1\leq j \leq l)$ is similar.
A cost model will be proposed later in Section \ref{sec:CostModelforPivotsSelection} to guide how to choose good social-network or road-network pivots.

\noindent{\bf Non-Leaf Nodes.}
Each entry $e$ of non-leaf nodes in index $\mathcal{I}$ is a {\it minimum bounding rectangle} (MBR) for all subgraphs under $e_{\mathcal{I}}$. In addition, $e$ is associated with a keyword super-set $e.key$ $(= \bigcup_{\forall u \in e} u.key)$ and $lb\_\Phi(e)$ $(= \min_{ \forall u \in e}\Phi(u))$. We maintain a bit vector $e.V_{key}$ for entry $e$ which is a bit-OR of bit vectors $u.V_{key}$ for all $u \in e$. In addition, we store a lower bound of edge support that is associated with each user under the node $e$ as follows: 
\begin{eqnarray}
 lb\_\Phi(e) = \min_{ \forall u \in e}\Phi(u).
 \label{eq:lb_w}
\end{eqnarray}

Finally, we store an upper bound of in-influence and out-influence scores, that is,\vspace{-2ex}

{\scriptsize
\begin{eqnarray}
\label{eq:node_Out_InfluenceScore}
   &&ub\_inf^{out}(e|\mathcal{T})\\ \nonumber 
   &=& \forall _{u\in e} \forall v \notin e, v\in u.deg_{out} \forall_{t \in \mathcal{T}} \{ \max\{tp^t_{u, v}\}, \dots, \max\{tp^{|\mathcal{T}|}_{u, v}\}\},
\end{eqnarray}
\begin{eqnarray}
\label{eq:node_IN_InfluenceScore}
   &&ub\_inf^{in}(e|\mathcal{T})\\ \nonumber 
   &=& \forall _{u\in e} \forall v \notin e, v \in u.deg_{in} \forall_{t \in \mathcal{T}} \{ \max\{tp^t_{v,u}\}, \dots, \max\{tp^{|\mathcal{T}|}_{v,u}\}\}.
\end{eqnarray}\vspace{-2ex}
}

We also store upper/lower bounds of actual road-network shortest path distance from each user's check-in locations to all road-network pivots $\mathcal{P}_{RN}$, and to social-network distances (the number of hops) to the social-network pivots $\mathcal{P}_{SN}$, that is, 
\begin{eqnarray}
 lb\_dist_{RN}(e, rpiv_k) = \min_{\forall u \in e}\{avg\_dist_{RN}(u, rpiv_k)\},
 \label{eq:lb_dist_{RN}NodeLevel}\\
 ub\_dist_{RN}(e_{\mathcal{I}}, rpiv_k) = \max_{\forall u \in e} \{avg\_dist_{RN}(u, rpiv_k)\},\label{eq:ub_dist_{RN}NodeLevel}\\
 lb\_dist_{SN}(e, spiv_k) = \min_{\forall u \in e} \{ dist_{SN}(u, spiv_k)\},\label{eq:lb_dist_{SN}NodeLevel}\\
 ub\_dist_{SN}(e, spiv_k) = \max_{\forall u \in e} \{ dist_{SN}(u, spiv_k)\}\label{eq:ub_dist_{SN}NodeLevel}.
\end{eqnarray}

\subsection{Index-Level Pruning}
In this subsection, we discuss the pruning on the social-spatial index $\mathcal{I}$ which can be used for filtering out (a group of) false alarms on the level of index nodes.

\noindent {\bf Spatial Distance-based Pruning for Index Nodes:} We utilize the road-network distance for ruling out index node $e_{\mathcal{I}_{i}}$ where users reside far away from locations of users in the candidate set $S$. Specifically, we have the following lemma.

\begin{lemma}(\textbf{Spatial Distance-based Pruning for Index Nodes}).
Given a spatial-social community $S$ of users from social network $G_s$, and a node $e_i$ from the social-spatial index $\mathcal{I}$. Node $e_i \in \mathcal{I}$ can be safely pruned, if $lb\_dist_{RN}(S, e_i) > \sigma$ holds, where $lb\_dist_{RN}(S, e_i)$ is the lower bound of the average road distance between users in the community $S$ and the index node $e_i$.  
\label{lem:SpatialDistance-basedPruningforIndexNodes}
\end{lemma}

\noindent {\it \uline{Discussion on Obtaining Lower Bounds of $dist_{RN}(S, e)$}}:
Next, we discuss how to derive the lower bound, $lb\_dist_{RN}(S, e)$, of the average road network distance which is used in Lemma \ref{lem:SpatialDistance-basedPruningforIndexNodes}. \vspace{-2ex}

{\scriptsize
\begin{eqnarray}
 && lb\_dist_{RN}(S, e)\\ \noindent
 &=& \max_{k=1}^l \left\{\begin{array}{ll}
& |dist_{RN}(u_q, rpiv_k) - lb\_dist_{RN}(e, rpiv_k)|,\\
& \qquad \mbox{\it if $dist_{RN}(u_q, rpiv_k) < lb\_dist_{RN}(e, rpiv_k)$;}\\
& |dist_{RN}(u_q, rpiv_k) - ub\_dist_{RN}(e, rpiv_k)|,\\ 
& \qquad \mbox{\it if $dist_{RN}(u_q, rpiv_k) > ub\_dist_{RN}(e, rpiv_k)$;}\\
& 0, \quad \mbox{\it otherwise,}\\
\end{array}
\right.\nonumber
\label{eq:lb_maxdist}
\end{eqnarray}
}where $u_q$ is the query vertex assigned at query time, and $lb\_dist_{RN}(e, rpiv_k)$ and $ub\_dist_{RN}(e, rpiv_k)$ are given in Eqs.~\eqref{eq:lb_dist_{RN}NodeLevel} and \eqref{eq:ub_dist_{RN}NodeLevel}, resp.

\noindent {\bf Influence Score Pruning for Index Nodes:}
The TCS-SSN query aims to produce communities where users highly influence each other.
For an ss-truss community $S$ and an index node $e \in \mathcal{I}$, node $e$ can be entirely pruned, if the influence score between the community $S$ and $e$ is less than threshold $\theta$.

\begin{lemma}({\bf Influence Score Pruning for Index Nodes}).
Given a spatial-social community $S$ and an index node $e \in \mathcal{I}$, $e$ can be safely pruned, if $lb\_infScore(S, e|\mathcal{T}) < \theta$ or $lb\_infScore(e, S|\mathcal{T}) < \theta$.
\label{lem:InfluenceScorePruningforIndexNodes}
\end{lemma}

\noindent {\it \uline{The Computatio of the Influence Score Lower Bound $lb\_infScore(S, e|\mathcal{T})$ on the Index $\mathcal{I}$}}:
We define the lower bound of the influence score between a spatial-social community $S$ and an index node $e \in \mathcal{I}$,with respect to the query vertex $u_q \in S$ as follows.
\begin{eqnarray}
lb\_infScore(S, e|\mathcal{T}_q)  =  ub\_inf^{out}(u_q|\mathcal{T}) \cdot ub\_inf^{in}(e|\mathcal{T}),
 \end{eqnarray}
 \begin{eqnarray}
lb\_infScore (e, S|\mathcal{T}_q)  =  ub\_inf^{out}(e|\mathcal{T}) \cdot ub\_inf^{in}(u_q|\mathcal{T}), 
\end{eqnarray}
where $ub\_inf^{in}(u_q|\mathcal{T})$ and $ub\_inf^{Out}(u_q|\mathcal{T})$ are given resp. in Eqs.~\eqref{eq:userIN_influenceScoreUperBound} and \eqref{eq:userOut_influenceScoreUperBound}, and  $ub\_inf^{in}(e|\mathcal{T})$ and $ub\_inf^{out}(e|\mathcal{T})$ are given in Eqs.~\eqref{eq:node_IN_InfluenceScore} and \eqref{eq:node_Out_InfluenceScore}, resp.

\noindent {\bf Social Distance-based Pruning for Index Nodes:}
An index node $e$ of index $\mathcal{I}$ can be filtered out by applying the social distance-based pruning, if the number of hops between the candidate community $S$ and users in $e$ is greater than a threshold $d$.

\begin{lemma}(\textbf{Social Distance-based Pruning for Index Nodes}).
Given a community $S$ of candidate users from social network $G_s$, and a node $e$ from index $\mathcal{I}$, a node $e \in \mathcal{I}$ can be safely pruned, if $lb\_dist_{SN}(S, e) > d$ holds, where $lb\_dist_{SN}(S, e)$ is the lower bound of the number of hops between users in $S$ and index node $e$.
\label{lem:SocialDistance-basedPruningforIndexNodes}
\end{lemma}

\noindent {\it \uline {Discussion on Obtaining Lower Bounds of $dist_{SN}(S, e)$}}:
Next, we discuss how to derive lower bound to derive the lower bound, $lb\_dist_{SN}(S, e)$, of the social-network distance (i.e., No. of hops) between the ss-truss $S$ and index node $e$.

To estimate the lower bound $lb\_dist_{SN}(S, e)$ of the social-network distance, we utilize social-network pivots as follows:

{\scriptsize
\begin{eqnarray}
 && lb\_dist_{SN}(S, e)\\ \noindent
 &=& \max_{k=1}^h \left\{\begin{array}{ll}
& |dist_{SN}(u_q, spiv_k) - lb\_dist_{SN}(e, spiv_k)|,\\
& \qquad \mbox{\it if $dist_{SN}(u_q, spiv_k) < lb\_dist_{SN}(e, spiv_k)$;}\\
& |dist_{SN}(u_q, spiv_k) - ub\_dist_{SN}(e, spiv_k)|,\\ 
& \qquad \mbox{\it if $dist_{SN}(u_q, spiv_k) > ub\_dist_{SN}(e, spiv_k)$;}\\
& 0, \quad \mbox{\it otherwise,}\\
\end{array}
\right.\nonumber
\label{eq:lb_maxdist}
\end{eqnarray}
}where $u_q$ is the query vertex assigned at query time, and $lb\_dist_{SN}(e, spiv_k)$ and $ub\_dist_{SN}(e, spiv_k)$ are offline pre-computed in Eqs.~\eqref{eq:lb_dist_{SN}NodeLevel} and \eqref{eq:ub_dist_{SN}NodeLevel}, respectively.

\noindent {\bf Structural Cohesiveness Pruning for Index Nodes.}
Similar to the structural cohesiveness pruning discussed in Section \ref{subsec:StructuralCohesivenessPruning}, if the lower  bound  of  edge  support  associated  with  users  under node $e \in \mathcal{I}$ is less than threshold $k$, then there is no edge under $e$ satisfying structural cohesiveness, and the node $e$ can be directly pruned. 

\begin{lemma}({\bf Structural Cohesiveness Pruning for Index Nodes})
Given a social-spatial index node $e\in \mathcal{I}$, if $lb\_\Phi(e)< k$, then the node $e$ can be safely filtered out.
\label{lem:StructuralCohesivenessPruningforIndex Nodes}
\end{lemma}

In Lemma \ref{lem:StructuralCohesivenessPruningforIndex Nodes}, $lb\_\Phi(e)$ is the lower bound of edge support of the index node $e$, defined in Section \ref{subsec:Social-SpatialIndex}. Intuitively, if all the edges associated with vertices under the node $e$ has a maximum support value that is less than $k$, then all vertices (users) under $e$ cannot be in the query result. The lower bound of edge support of the node $e$ is computed in Eq.~\eqref{eq:lb_w}.

\noindent {\bf Keyword-based Pruning for Index Nodes:}
Definition \ref{def:TCS-SSN}, ensures that each user in the returned community contains at least one keyword query that appears in the query set $K_q$. Therefore, for an index node $e \in \mathcal{I}$ can be safely pruned, if all users under $e$ share no keywords with the keyword query set $K_q$.

 \begin{lemma}({\bf Keyword-based Pruning for Index Nodes})
 Given an index node $e \in \mathcal{I}$ and a set $K_q$ of query keywords, node $e$ can be safely ruled out, if $e.V_{key} \cap K_q = \emptyset$.
 \label{lem:Keyword-basedPruningforIndexNodes}
 \end{lemma}

For an index node $e$, if it holds that $e.V_{key} \cap K_q = \emptyset$, it indicates that node $e$ does not contain any keywords in $K_q$, and thus $e$ can be pruned.

\subsection{The Construction of a Social-Spatial Index}
\label{sec:BuildingtheSocial-SpatialIndex}
 Algorithms \ref{alg:piv_selection} and \ref{alg:FindSubgraphs} will be running simultaneously to generate the social-spatial index $\mathcal{I}$.
The general idea of building the social-spatial index is to; First, find a number $\iota$ of index pivots (social network users); Second, partition the social network users (vertices) around those pivots.

We first start by describing Algorithm \ref{alg:FindSubgraphs}, where the input is a social network $G_s$, a spatial network $G_r$, and a set $\mathcal{P}_{index}$ of $\iota$ pivots (social-network vertices). The goal is to generate $\iota$ subgraphs around $\mathcal{P}_{index}$.

For each social-network vertex $v$, we compute the quality with each social-network pivot $piv_i \in \mathcal{P}_{index}$ (lines 1-4). The quality function $quality(v,piv_i)$ computes the number of hops and road-network distance between $v$ and $piv_i$ (line 4). Then, assign the vertex $v$ to the pivot where the quality is the best (lines 5-8). Finally, the set of partitions in returned (line 9). 

Algorithm \ref{alg:piv_selection}, illustrates the details of the pivot selection. At the beginning, two parameters $global_{cost}$ and $\mathcal{P}$ will be set to store the globally optimal cost value and the corresponding pivot set, resp. (line 1).
We randomly select a pivot set $\mathcal{S}_p$ from social-network users (vertices) (line 3). Next, we partition the social network around $\mathcal{S}_p$ by Algorithm \ref{alg:FindSubgraphs} and partitions $\mathcal{G}$ (line 4). Then, we evaluate the cost function $Cost\_\mathcal{P}_{index}(\mathcal{G})$ of the resulting partitions by Eq.~\eqref{eq:costModel} (line 5). After that, each time we swap a $piv\in \mathcal{S}_p$ with a non-pivot $new\_piv$, which results in a new pivot set $\mathcal{S}_p'$ (lines 7-9), and generate new graph partitions $\mathcal{G'}$ by Algorithm \ref{alg:FindSubgraphs} and evaluate it (lines 10-11). If the new cost is better than the best-so-far cost $local\_cost$, then we can accept the new pivot set with its cost (lines 12-14).
We repeat the process of swapping a pivot with a non-pivot for $swap\_iter$ times (line 6).
To avoid the local optimal solution, we consider selecting different initial pivot sets for $globa\_liter$ times (lines 2-3), and record the globally optimal pivot set and its cost (lines 15-17). Finally, we return the best pivot set $\mathcal{P}_{index}$.

Finally, we pass the optimal pivot set $\mathcal{P}_{index}$ to Algorithm \ref{alg:FindSubgraphs} to generate subgraphs, which are treated as leafs of the social-spatial index. Then, the connected subgraphs in leaf nodes are recursively grouped into non-leaf nodes, until a final root is obtained.

\subsection{The Evaluation Measure of Social-Spatial Index $\mathcal{I}$}
\label{subsec:DesigningtheSocial-SpatialIndex}
We design our index to group potential user communities together. The criteria of the grouping are the spatial distance, structural cohesiveness, and the influence score.
We use these three criteria to measure the quality of the formed subgraphs.
Our goal is to group social-network users who are spatially close, having small social distance (i.e., the number of hops), having high structural cohesiveness, and mutually influences each other in one group or neighbouring groups.  
We consider three factors to evaluate the quality of the produced subgraphs, that is {\it spatial closeness, structural cohesiveness}, and {\it social influence}.

\noindent \textit{\underline{Spatial Closeness}}:
The spatial closeness of social-network users in subgraph $g$ of $G_s$ is given by function $\chi_{sc}$ as follows.
\begin{eqnarray}
 \chi_{sc}=\sum_{\forall g\in G_s} \sum_{\forall u_j\in V(g)}\sum_{\forall u_k\in V(g)} avg\_dist_{RN}(u_j, v_k).
 \label{eq:SpatialClosenessMeasure}
\end{eqnarray}

Since each social-network user may have multiple check-in locations, we utilize Eq. (\ref{eq:spatialShortestPath}) to evaluate the shortest path distance between two users. As an example in Figure \ref{fig:ssn}, if we form a social spatial group for $u_5$ based on the spatial distance, at first $u_1, u_2, u_3, u_4$, and $u_6$ are candidates. 
Form the spatial network, since $u_4$ and $u_6$ are spatially close to $u_5$, $u_5$ is most likely to form a group with them.
Eq. (\ref{eq:SpatialClosenessMeasure}) ensures that two far vertices such as $u_5$ and $u_1$ can be distributed to two different subgraphs.

\noindent \textit{\underline{Structural Cohesiveness}}:
The structural closeness $\chi_{st}$ measures structural cohesiveness and social-network distance among users of subgraph $g \in G_s$ as follows: 

\begin{eqnarray}
 \chi_{st}=\sum_{\forall g\in G_s} \sum_{\forall u_j\in V(g)}\sum_{\forall u_k\in V(g)} 
 \frac{\Phi(u_j) + \Phi(u_k)}{Dist_{SN}(u_j, u_k)}, 
 \label{eq:StructuralCohesivenessMeasure}
\end{eqnarray}

  Intuitively, in Eq.~(\ref{eq:StructuralCohesivenessMeasure}) social-network users who have high structural cohesiveness and small social-network distance will be in the same subgraph or neighboring subgraphs. 

\noindent \textit{\underline{Social Influence}}:
In social networks, users influence each other based on topics they like. We use the influence score function in Eq.~\eqref{eq:infScore_TwoVertices} to measure the influence of two users in subgraphs of $G_s$.
\begin{eqnarray}
 \chi_{inf}=\sum_{\forall g\in G_s} \sum_{\forall u_j\in V(g)}\sum_{\forall u_k\in V(g)} infScore(u_j, u_k), 
 \label{eq:SocialInfluenceMeasure}
\end{eqnarray}

To implement social influence, in our social-spatial index $\mathcal{I}$, we gather social-network users who highly influence each other in the same subgraph. In Eq.~\ref{eq:SocialInfluenceMeasure} social-network users who highly influence each other will be gathered within subgraphs.

\subsection{Cost Model for the Pivot Selection}
\label{sec:CostModelforPivotsSelection}
In this subsection, we discuss our algorithms of selecting good social-network pivots $\mathcal{P}_{SN}$, road-network pivots $\mathcal{P}_{RN}$, and index pivots $\mathcal{P}_{index}$.
We utilize the social-network pivots $\mathcal{P}_{SN}$ for the social distance-based pruning in Section \ref{subseq:SocialDistance-BasedPruning}. Similarly, the road network pivots $\mathcal{P}_{RN}$ are used for the spatial distance based pruning in Section \ref{subseq:SpatialDistance-BasedPruning}. The index pivots $\mathcal{P}_{index}$ are employed in building the social-spatial index $\mathcal{I}$, as mentioned in Section \ref{sec:BuildingtheSocial-SpatialIndex}.

\subsubsection{Cost Model for the Road-Network Pivots, $\mathcal{P}_{RN}$, Selection}
As discussed in Section \ref{subseq:SpatialDistance-BasedPruning}, we aim to utilize the road-network pivots $\mathcal{P}_{RN}$ to derive the upper bound of the average spatial distance between any two social-network users $u$ and $v \in G_s$.
We filter out false alarms of user nodes $e$ (or objects) that are far away from users in $S$. Essentially, the pruning power of this method depends on the tightness of lower bound of the average distance (derived via pivots) between user $u.L \in G_s$ and user $v.L$. Therefore, we define a cost function, $Cost_{\mathcal{P}_{RN}}$, as the difference (tightness) of lower distance bounds via pivots, which can be used as a measure to evaluate the goodness of the selected road-network pivots.

In particular, we have the following function:
\begin{eqnarray}
\label{eq:cost_RnPivots}
&& Cost\_{\mathcal{P}_{RN}} \\ \nonumber
&=& \hspace{-3ex}\sum_{\forall u, u \in V(G_s)} \max_{k=1}^{l}|avg\_dist_{RN}(u,rpiv_k)-avg\_dist_{RN}(v,rpiv_k)|.\hspace{-5ex}\nonumber\\\nonumber
\end{eqnarray}
Our goal is to select road-network pivots $\mathcal{P}_{RN} = \{ rpiv_1, \dots, rpiv_l\}$ that minimize the cost model $Cost\_{\mathcal{P}_{RN}}$ (given in Eq.~\eqref{eq:cost_RnPivots}).

\subsubsection{Cost Model for the Social-Network Pivots, $\mathcal{P}_{SN}$, Selection}
As mentioned in Section \ref{subseq:SocialDistance-BasedPruning}, the social distance pruning
rules out false alarms of user $u$ with distance lower bound $lb\_dist_{SN}(u, v)$ greater than or equal to threshold $d$. 
The distance lower bound can be computed via $h$ pivots (users), $spiv_k$, that is, $lb\_dist_{SN}(u, v) = |dist_{SN}(u, spiv_k) - dist_{SN}(v, spiv_k)|$. Intuitively, larger distance lower bound leads to higher pruning power.

\begin{eqnarray}
\label{eq:cost_SnPivots}
 &&Cost\_{\mathcal{P}_{SN}} \\ \nonumber 
 &=&\hspace{-3ex}\sum_{\forall u, u \in  V(G_s)}\max_{k=1}^{h}|dist_{SN}(u,spiv_k)-dist_{SN}(v,spiv_k)|
\end{eqnarray}

Thus, our target is to choose good social-network pivots $\mathcal{P}_{SN}$ that maximize the cost $Cost_{\mathcal{P}_{SN}}$ (given in Eq.~\eqref{eq:cost_SnPivots}).

\subsubsection{Cost Model for the Index Pivots, $\mathcal{P}_{index}$, Selection}
 As explained in Section \ref{subsec:DesigningtheSocial-SpatialIndex}, our social-spatial index groups potential communities together. Algorithm \ref{alg:FindSubgraphs} utilizes the index pivots $\mathcal{P}_{index}$ to build communities around those pivots. 
 In Section \ref{subsec:DesigningtheSocial-SpatialIndex}, we developed three measures spatial closeness measure, structural cohesiveness, and social influence measure.
 
 Next, develop a cost model function $Cost\_{\mathcal{P}_{index}}(G_s)$ by evaluating the produced subgraphs resulting from partitioning with pivots $\mathcal{P}_{index}$ as follows:

\begin{eqnarray}
 \label{eq:costModel}
 &&Cost\_{\mathcal{P}_{index}}(G_s)\\
 &=& \mathcal{W}_1 \cdot \chi_{sc} + \mathcal{W}_2 \cdot (1- \chi_{st}) + \mathcal{W}_3 \cdot (1- \chi_{inf}).\nonumber
\end{eqnarray}

Our goal is to select social-network pivots $\mathcal{P}_{index}$ that maximize the cost function $Cost\_{\mathcal{P}_{index}}(G_s)$ (given in Eq.~\eqref{eq:costModel}).

\begin{algorithm}[t!]\scriptsize
    \KwIn{a road network $G_r$, a social network $G_s$, and the number $\iota$ of pivots}
    \KwOut{the set, $\mathcal{P}_{index}$, of pivots}
    $global\_{cost}=-\infty$, $\mathcal{P}=\emptyset$;\\
    \For{$a = 1$ to $global\_iter$}
    {
        randomly select $\iota$ initial pivots and form a pivot set $\mathcal{S}_{p}$\\
        generate subgraphs based on pivots $\mathcal{G}$ = {\sf Gen\_Subgraphs}$(G_r, G_s,\mathcal{S}_{p})$\\
        set $local\_cost = Cost\_{\mathcal{P}_{index}}({\mathcal{G}})$\\
        \For{$b = 1$ to $swap\_iter$}{
           select a random pivot $piv \in \mathcal{S}_{p}$\\ 
           randomly choose a non-pivot $new\_piv$\\
           $\mathcal{S}_{p}^{'}= \mathcal{S}_{p}-\{piv\}+\{new\_piv\}$\\
           $\mathcal{G}^{'}$ = {\sf Gen\_Subgraphs}$(G_r, G_s,\mathcal{S}_{p}^{'})$\\
           evaluate the new cost $Cost\_{\mathcal{P}_{index}}^{new} ({\mathcal{G}}{'})$ w.r.t. $\mathcal{S}_{p}^{'}$\\
           \If {the new cost $Cost\_{\mathcal{P}_{index}}^{new} ({\mathcal{G}}{'})$ is better than $local\_cost$}{
                $local\_cost = Cost\_{\mathcal{P}_{index}}^{new} ({\mathcal{G}}{'})$\\
                $\mathcal{S}_{p}=\mathcal{S}_{p}^{'}$
            }
        }
        \If{$local\_cost$ is better than $global\_cost$}{
            $\mathcal{P}_{index}=\mathcal{S}_{p}$\\
            $global\_cost=local\_cost$
        }
    }
 \Return $\mathcal{P}_{index}$
    \caption{\sf Index Pivot Selection}
    \label{alg:piv_selection}\vspace{3ex}
\end{algorithm}

\begin{algorithm}[t!]\scriptsize
    \KwIn{a spatial network $G_r$, a social network $G_s$, and a set $\mathcal{P}_{index}= piv_1, \dots, piv_h$ of pivots}
    \KwOut{a set $\mathcal{G}= {g_1, \dots, g_\iota}$ of subgraphs}
    {
        \For{$v \in V(G_s)$}{
            $best\_quality = \infty$\\
            \For{$i= 1$ to $\iota$}{
                $quality(v, piv_i)= \dfrac{avg\_dist_{RN}(v, piv_i)}{max\_avg\_dist_{RN}} + \dfrac{dist_{SN}(v, piv_i)}{max\_Dist_{SN}}$\\
                \If {$quality(v, piv_i) < best\_quality$}{
                    $j = i$\\
                    $best\_quality = quality(v, piv_i)$
                }
            }
            assign $v$ to $g_j$
        }
    }
 \Return $\mathcal{G}$
    \caption{\sf Gen\_Subgraphs}
    \label{alg:FindSubgraphs}\vspace{3ex}
\end{algorithm}

\section{Community Search Query Answering}
\label{sec:Query_Answering}
Algorithm \ref{alg:TCS-SSN_processing} illustrates the pseudo code of {\sf TCS-SSN} answering, which process TCS-SSN queries over the spatial-social network $G_{rs}$ via the social-spatial index $\mathcal{I}$.
Specifically, we traverse index $\mathcal{I}$, and apply index level pruning over the index node and objects level pruning over the social network object, and refine a candidate set to return the actual TCS-SSN query answer.

\noindent {\bf Pre-Processing.} Initially, we set $S_{cand}$ to an empty set, initialize an empty minimum heap $\mathcal{H}$, and add the root, $root(\mathcal{I})$, of index $\mathcal{I}$ to $\mathcal{H}$ (lines 1-3).

\noindent {\bf Index Traversal.}
In Algorithm \ref{alg:TCS-SSN_processing}, after we insert the heap entry $(root(\mathcal{I}), 0)$ into the heap $\mathcal{H}$, we traverse the social-spatial index $\mathcal{I}$ from root to leaf nodes (lines 4-15). 
In particular, we will use heap $\mathcal{H}$ to enable the tree traversal. Each time we pop out an entry $(e_i, key)$ with the minimum key from heap $\mathcal{H}$, where $e_i$ is an index node $e_i\in \mathcal{I}$, and $key$ is a lower bound of road-network distance, $key = lb\_dist_{RN}(e, e_i)$. If $key$ is greater than spatial distance threshold $\sigma$, all entries in $\mathcal{H}$ must have their lower bounds of maximum road-network distances greater than threshold $\sigma$. Then, we can safely prune all entries in the heap and terminate the loop. 

When entry $e_i$ is a leaf node, we consider each object (social-network user) $u \in e_i$, and apply object-level pruning {\it spatial distance-based pruning, influence score pruning, structural cohesiveness pruning, social distance-based pruning,} and {\it keyword-based pruning} to reduce the search space (line 9). If a user $u$ cannot be pruned, we will add it to the candidate set $\mathcal{S}_{cand}$ (line 10).

When entry $e_i$ is a non-leaf node, for each child $e_x \in e_i$, we will apply index-level pruning (e.g., spatial distance-based pruning influence score pruning, social distance-based pruning, structural cohesiveness pruning, and keyword-based pruning for index nodes) (line 14). If a node $e_x$ cannot be pruned in line 14, then we insert heap entry ($e_i, lb\_dist_{RN}(q, e_x)$) into heap $\mathcal{H}$ for further investigation (line 15).

\noindent {\bf Refinement.} After the index traversal, we refine the candidate set $\mathcal{S}_{cand}$ to obtain/return actual TCS-SSN answers $\mathcal{S}$ (line 17).

\noindent{\bf Complexity Analysis.}
Next, we discuss the time complexity of our TCS-SSN query answering algorithm in Algorithm \ref{alg:TCS-SSN_processing}.
The time cost of Algorithm \ref{alg:TCS-SSN_processing} processing consists of two portions: index traversal (lines 4-15) and refinement (lines 16-18).

Let $PP^{(j)}$ be the pruning power on the $j$-th level of index $\mathcal{I}$, where $1 \leq j \leq height(\mathcal{I})$. Denote $f$ as the average fanout of non-leaf nodes in the social-spatial index $\mathcal{I}$.
Then, the filtering cost of lines 4-15 is given by $O\big(\sum_{j=1}^{height(I)} f^j\cdot (1 - PP^{(j-1)})\big)$, where $PP^{(0)} = 0$.

Moreover, let $S_{cand}$ be a subgraph containing users left after applying our pruning methods.
The main refinement cost in lines 16-18 is on the graph traversal and constraint checking (e.g., average spatial distance, social distance, and social influence). 
In particular, the average spatial distance on road networks can be computed by running the Dijkstra algorithm starting from every vertex in $S_{cand}$, which takes $O(|V_{RN}(S_{cand})| \cdot (|E_{RN}(S_{cand})| \cdot log(|V_{RN}(S_{cand})|)))$ cost;
the social distance computation takes $O(|V_{SN}(S_{cand})| \cdot |E_{SN}(S_{cand})|)$ by BFS traversal from each user in $S_{cand}$; the k-truss computation takes $O(p \cdot |E_{SN}(S_{cand})|)$, where $p < {min(d_{max}, \sqrt{|E_{SN}(S_{cand})|}})$ \cite{huang2017attribute}; 
the mutual influence score computation takes $O(|V_{SN}(S_{cand})| \cdot |E_{SN}(S_{cand})|)$ by BFS traversal from each user in $S_{cand}$. Thus, the overall time complexity of the refinement is given by $O( |V_{RN}(S_{cand})| \cdot (|E_{RN}(S_{cand})| \cdot log(|V_{RN}(S_{cand})|)) + p \cdot |E_{SN}(S_{cand})| + 2 \cdot (|V_{SN}(S_{cand})| \cdot |E_{SN}(S_{cand})|))$.

\begin{algorithm}[t!]\scriptsize
    \KwIn{a spatial-social network $G_{rs}$, social-spatial index $\mathcal{I}$, a query issuer $q$, 
    a topic query set $\mathcal{T}_q$, a keyword query set $K_q$, a truss value $k$, social distance threshold $d$, spatial distance threshold $\sigma$, influence threshold $\theta$}
    \KwOut{a community $S$, satisfying TCS-SSN query predicates in Definition \ref{def:TCS-SSN}}
    
    set $S\_{cand} = \emptyset$\\
    initialize a min-heap $\mathcal{H}$ accepting entries in the form $(e, key)$\\
    insert entry $(root(\mathcal{I}), 0)$ into heap $\mathcal{H}$\\
        \While{$\mathcal{H}$ is not empty}{
            $(e_i, key)$ = de-heap $\mathcal{H}$\\
            {\bf if} $key > \sigma$, {\bf then} terminate the loop;\\
            \If {$e_i$ is a leaf node}{
                \For {each user $u\in e_i$}{
                    \If {$u$ cannot be pruned by Lemma \ref{lem:UpperBoundoftheAverageSpatialDistance},  \ref{lem:UB_InfluenceScorePruning},  \ref{lem:StructuralCohesivenessPruning},  \ref{lem:SocialDistance-BasedPruning}, or \ref{lem:Keyword-basedPruning} w.r.t. $q$}{
                            add $u$ to $S_{cand}$ 
                    }
                }
            }
            \Else{ \tcp{$e_i$ is a non-leaf node}
                obtain the entry $e_q \in \mathcal{I}$ that contains $q$\\
                \For {each entry $e_x\in e_i$}{
                    \If {$e_x$ cannot be pruned by Lemma \ref{lem:SpatialDistance-basedPruningforIndexNodes},  \ref{lem:InfluenceScorePruningforIndexNodes}, \ref{lem:SocialDistance-basedPruningforIndexNodes},  \ref{lem:StructuralCohesivenessPruningforIndex Nodes}, or
                     \ref{lem:Keyword-basedPruningforIndexNodes} w.r.t. $e_q$}{
                            insert $(e_x, lb\_dist_{RN}(q, e_x))$ into the heap $\mathcal{H}$
                }
            }
        }
    }
    \While{no users or edges are pruned}{
    start BFS search from $q$ and apply social-network distance pruning and influence score pruning on social network vertices \\
    apply truss decomposition directly on remaining edges to prune edges with truss value less than or equal to $k-2$
    }

    \Return $S$
    \caption{\sf TCS-SSN Query Answering Algorithm}
    \label{alg:TCS-SSN_processing}\vspace{3ex}
\end{algorithm}

\noindent {\bf Discussions on Handling Multiple Query Users.} The TCS-SSN problem considers the standalone community search issued by one query user $q$. In the case where multiple users issue the TCS-SSN queries at the same time, we perform batch processing of multiple TCS-SSN queries, by traversing the social-spatial index only once (applying our pruning methods) and retrieving candidate users for each query user. In particular, an index node can be safely pruned, if for each query there exists at least one pruning rule that can prune this node. After the index traversal, we refine the resulting candidate users for each query and return the TCS-SSN answer sets to query issuers.

\section{Experimental Evaluation}
\label{sec:ExperimentalEvaluation}

\begin{table}[t!]
\caption{\small Statistics of real data sets $Gow\&Cali$ and $Twi\&SF$.}
\label{table:realData}
{\scriptsize
\begin{tabular}{||p{1cm}|c|c||p{1.4cm}|c|c||}\hline
    {\bf \text{ } social} & {\bf $|V(G_s)|$} & {\bf $|E(G_s)|$}& {\bf \text{ } road} & {\bf $|V(G_r)|$} & {\bf $|E(G_r)|$}\\
    {\bf network} &  & & {\bf network} &  & \\\hline\hline 
    
    Gowalla ($Gow$) &196K & 1.9M & California ($Cali$) & 21K & 44K\\\hline\hline
    
    Twitter ($Twi$) &349K & 2.1M & San Francisco ($SF$) & 175K& 446K\\\hline

\end{tabular}
}
\end{table}

\begin{table}[t!]
\centering{\scriptsize
\caption{\small Experimental settings.}
\label{table:parameter}
\begin{tabular}{||p{3.8cm}|p{4cm}||}\hline
    {\bf Parameter} & {\bf Values} \\ \hline\hline
    the size of keyword set query $K_q$ & 2, 3, {\bf 5}, 8, 10\\\hline
    the size of topic set query $\mathcal{T}_q$ & 1, {\bf 2}, 3\\\hline    
    the spatial distance threshold $\sigma$ & 0.5, 1, {\bf 2}, 3, 5 \\\hline
    the influence score threshold $\theta$ & 0.1, 0.3, {\bf 0.5}, 0.7, 0.9 \\\hline
     the number of triangles $k$ & 2, 3, {\bf 5}, 7, 10 \\\hline
     the social distance threshold $d$ & 1,  2, {\bf 3}, 5, 10 \\\hline
    the number of vertices in road network $G_r$ and social network $G_s$ & 10K, 20K, {\bf 30K}, 40K, 50K, 100K , 200K \\\hline
\end{tabular}
}
\end{table}

\subsection{Experimental Settings}
We test the performance of our TCS-SSN query processing approach (i.e., Algorithm  \ref{alg:TCS-SSN_processing}) on both real and synthetic data sets.

\noindent {\bf Real Data Sets.} We evaluate the performance of our proposed TCS-SSN algorithm (as given in Algorithm  \ref{alg:TCS-SSN_processing}) with two real data sets, denoted as $Gow\&Cali$ and $Twi\&SF$, for spatial-social networks. The first data set, $Gow\&Cali$, is a spatial-social network, which combines Gowalla social network \cite{li2014efficient} with California road networks \cite{li2005trip}. The second spatial-social network $Twi\&SF$ integrates the Twitter \cite{li2014efficient} with San Francisco road networks \cite{li2005trip}. Table \ref{table:realData} depicts statistics of spatial/social networks.

Each user $u$ in social networks (i.e., Gowalla or Twitter) is associated with multiple check-in locations (i.e., places visited by the user $u$). The user $u$ also has a keyword vector $u.key$, which contains keywords collected from one's social-media profile. The directed edge $e_{u, v}$ between users $u$ and $v$ has a weight that reflects the influence of user $u$ on another user $v$ based on a certain topic. We map each user $u$ from social networks (Gowalla or Twitter) to 2D locations on road networks (i.e., California or San Francisco, respectively). 

\noindent {\bf Synthetic Data Sets.} We also generate two synthetic spatial-social data sets as follows. Specifically, for the spatial network $G_r$, we first produce random vertices in the 2D data space following either Uniform or Gaussian distribution. Then, we randomly connect vertices nearby through edges, such that all vertices are reachable in one single connected graph and the average degree of vertices is within $[3, 4]$. This way, we can obtain two types of graphs with Uniform and Gaussian distributions of vertices.

To generate a social network $G_s$, we randomly connect each user $u$ with other users, such that the degrees of users follow Uniform or Gaussian distribution within a range $[1, 10]$. Each user $u$ has a set, $u.key$, of interested keywords, where keywords are represented by integers within $[1, 10]$ following Uniform or Gaussian distribution. Furthermore, each social-network edge $e_{u, v}$ is associated with a set of topics (we consider 3 topics by default), and each topic has a probability (within $[0, 1]$ following Uniform or Gaussian distribution) that user $u$ can influence user $v$, similar to \cite{chen2015online}.

Finally, we combine social network $G_s$ with road network $G_r$, by randomly mapping social-network users to 2D spatial locations on road networks, and obtain a spatial-social network $G_{rs}$. With Uniform or Gaussian distributions during the data generation above, we can  obtain two types of synthetic spatial-social networks $G_{rs}$, denoted as $Uni$ and $Gau$, respectively.

\begin{figure}[t!]\hspace{-4ex}
\subfigure[][{\small CPU time}]{
\scalebox{0.25}[0.25]{\includegraphics{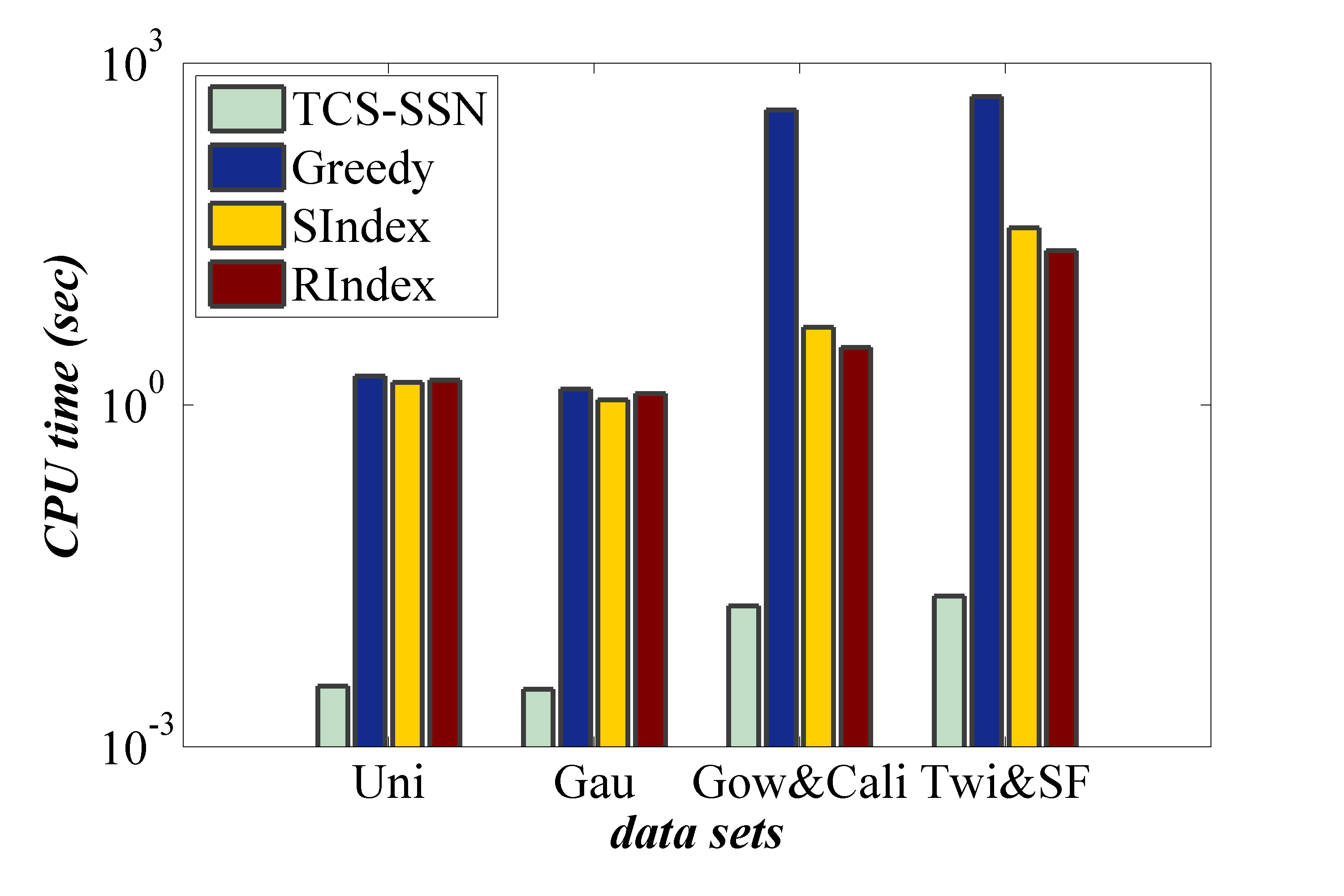}}\label{subfig:vs_time}
}%
\subfigure[][{\small I/O cost}]{\hspace{-3ex}    
\scalebox{.25}[.25]{\includegraphics{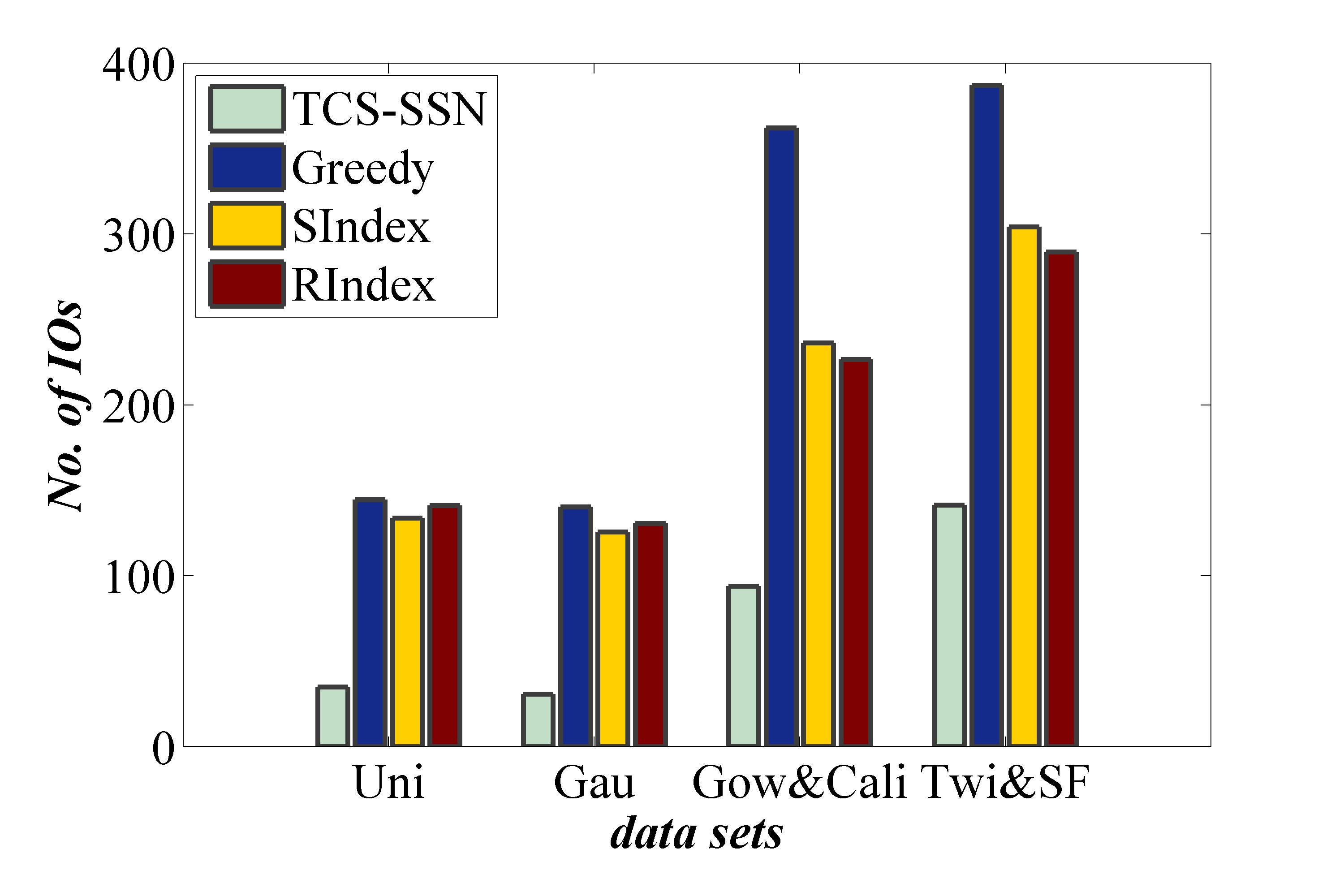}}\label{subfig:vs_IO}
}\vspace{-1ex}  
     \caption{\small The TCS-SSN performance vs. real/synthetic data sets.}
     \label{fig:vs}
\end{figure}

\begin{figure}[t]
\centering
\scalebox{.27}[.27]{\includegraphics{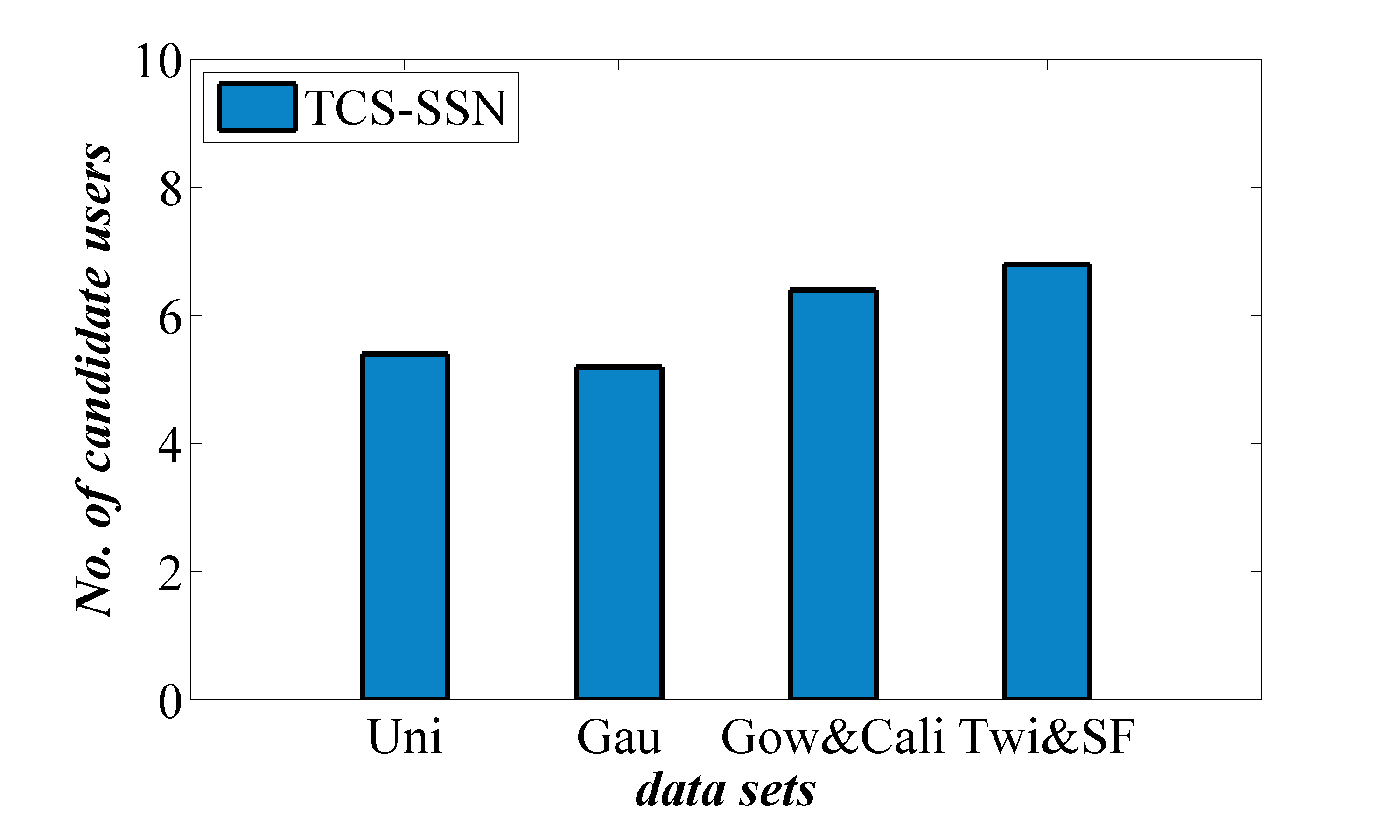}}
\caption{\small The number of the remaining candidate users after the pruning vs. real/synthetic data sets.}
\label{fig:candSize}
\end{figure}

\begin{figure}[t!]\hspace{-4ex}
\subfigure[][{\small index construction time}]{                    
\scalebox{0.25}[0.27]{\includegraphics{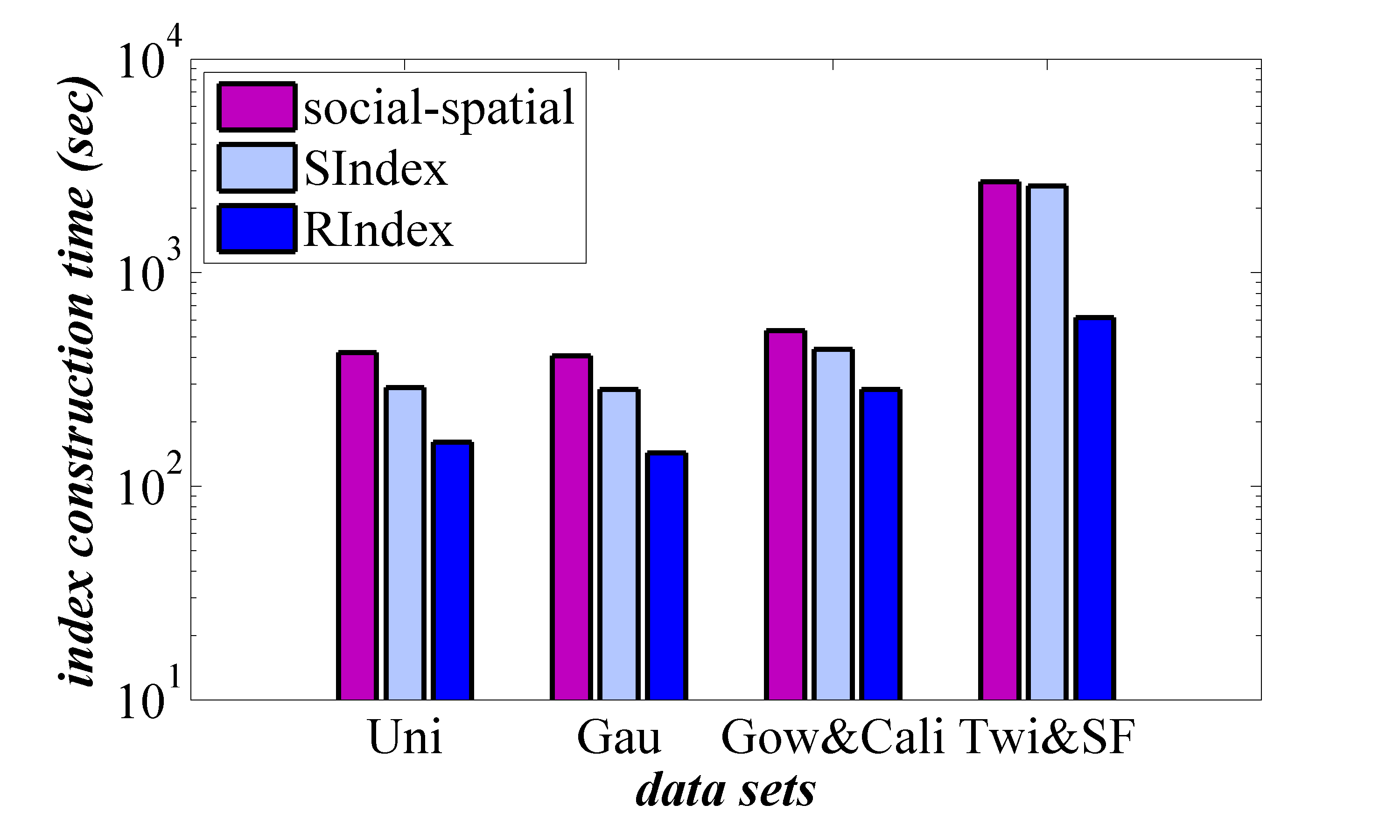}}\label{subfig:indexTime}
}%
\subfigure[][{\small index space cost}]{                    
\scalebox{.25}[.27]{\includegraphics{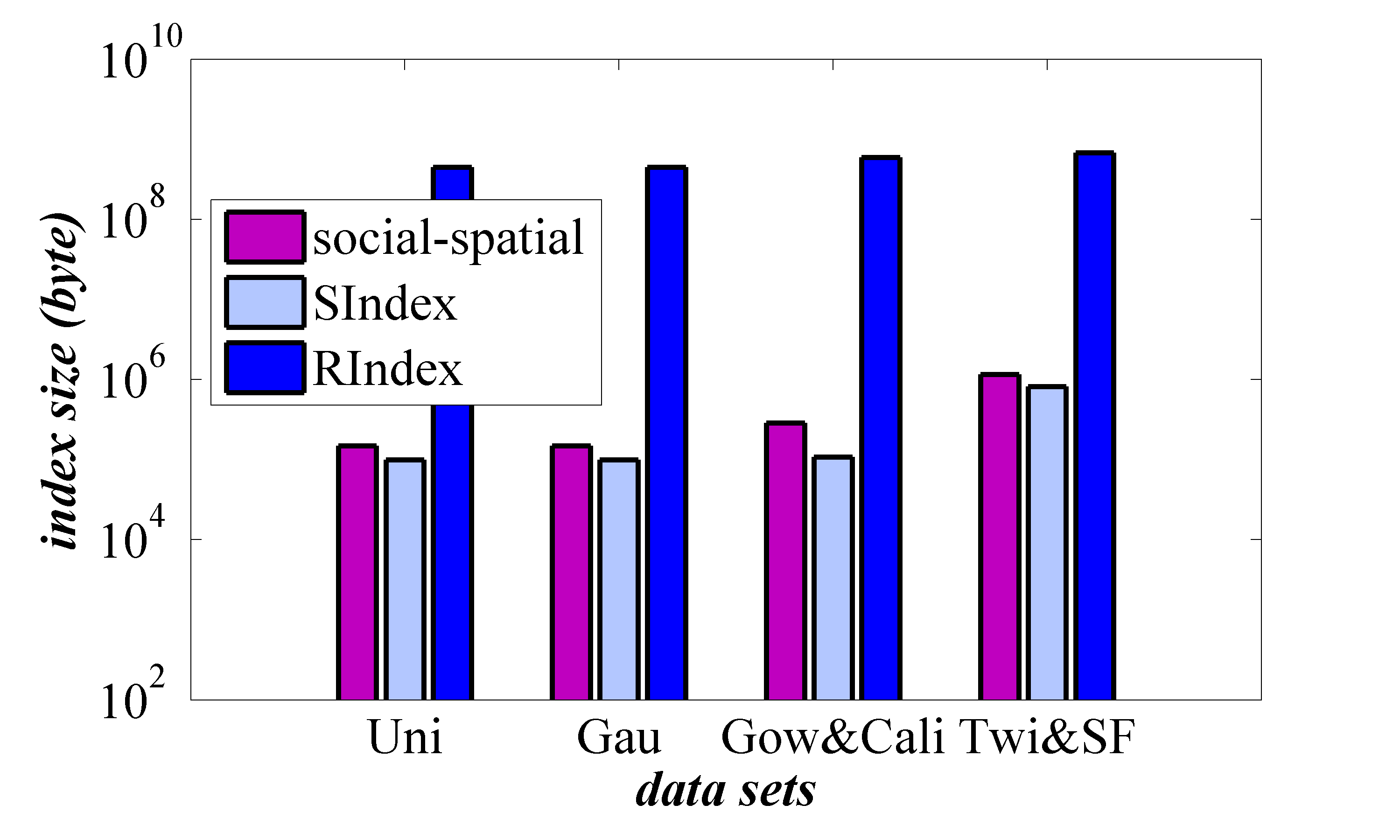}}\label{subfig:indexUp}
}\vspace{-1ex}  
     \caption{\small The index construction time and space cost vs. real/synthetic data sets.}
     \label{fig:index_time}
\end{figure}

\begin{figure}[t!]\hspace{-4ex}
\subfigure[][{\small CPU time}]{                    
\scalebox{0.3}[0.3]{\includegraphics{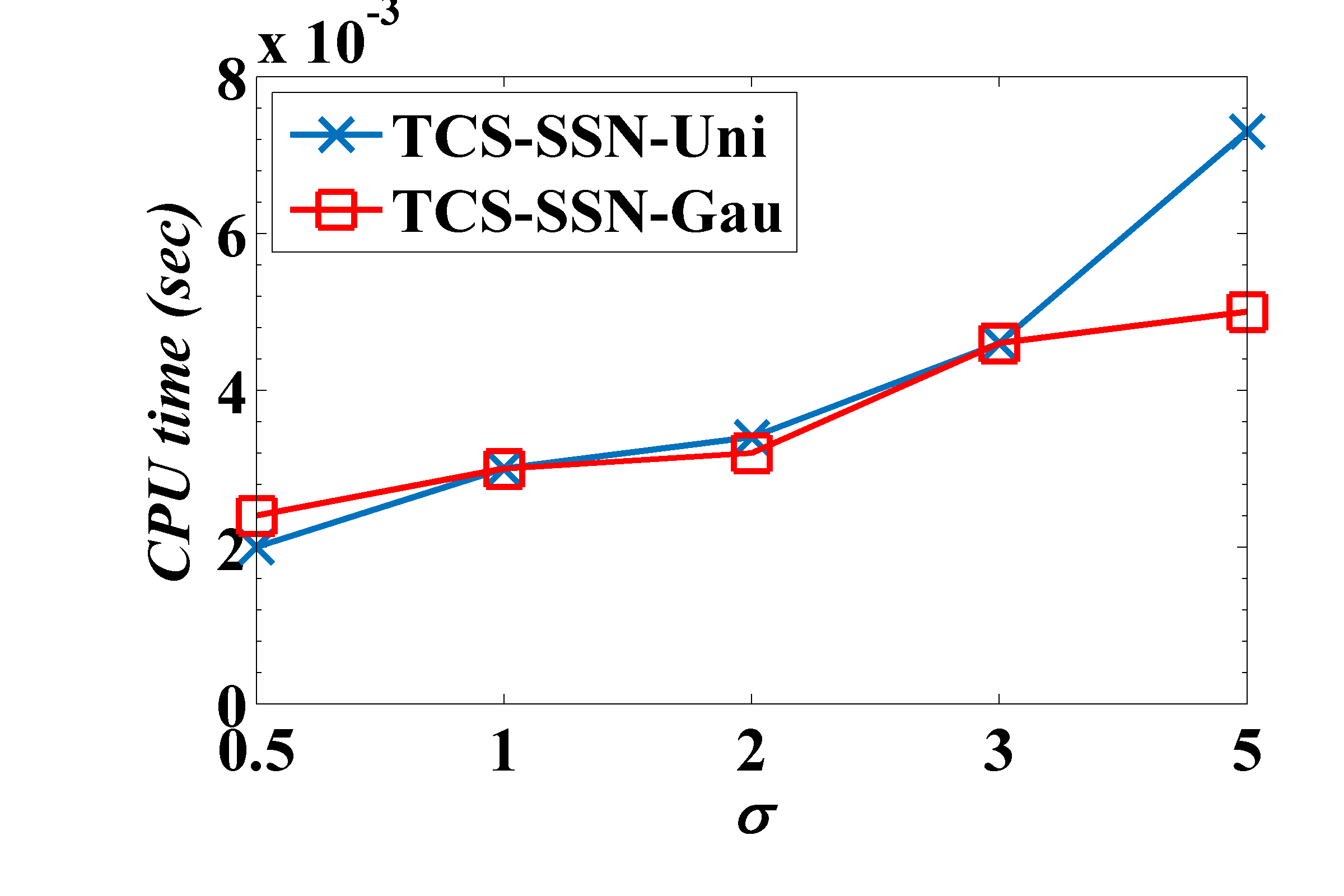}}\label{subfig:sigma_time}
}%
\subfigure[][{\small I/O cost}]{                    
\scalebox{.3}[.3]{\includegraphics{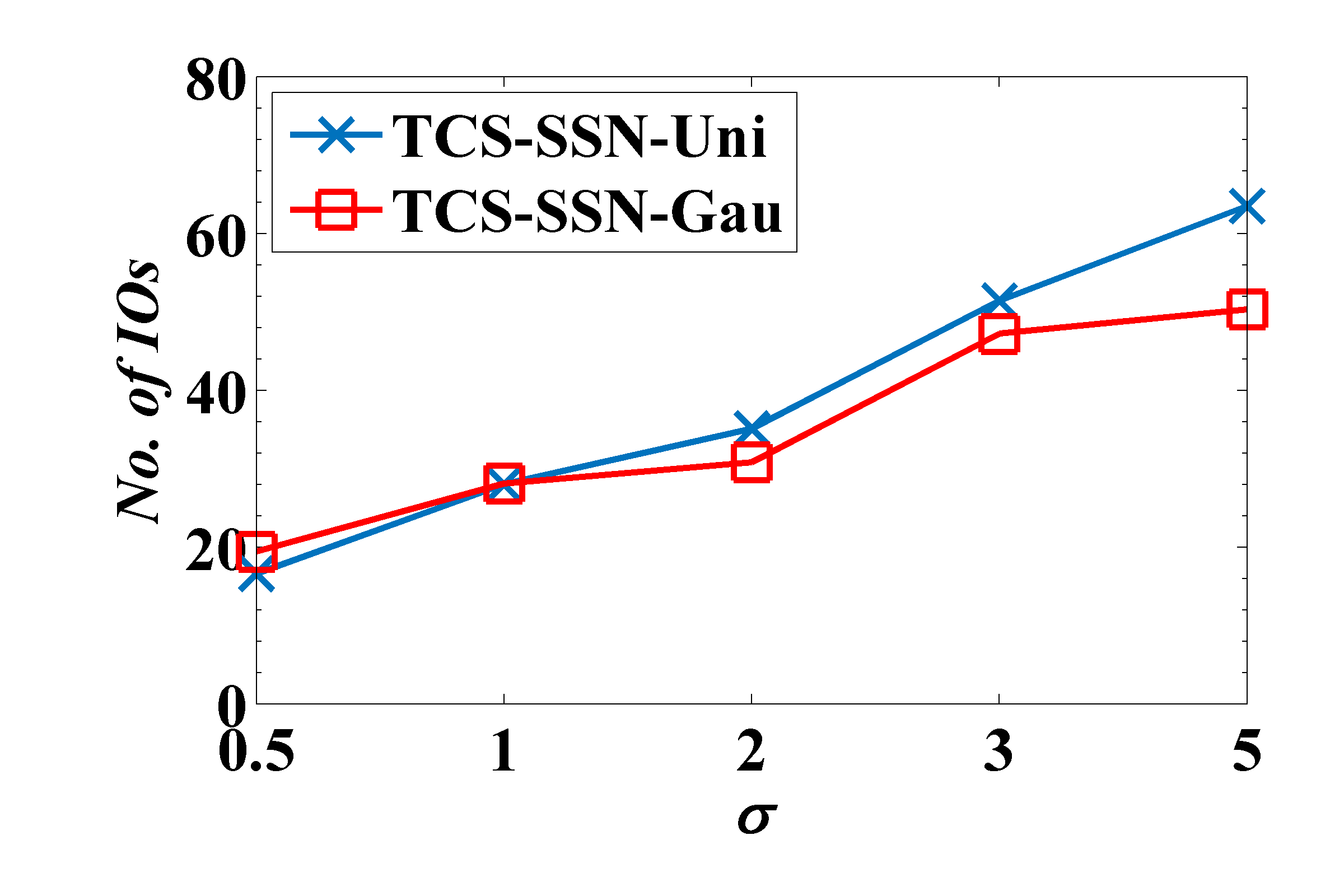}}\label{subfig:sigma_IO}
}\vspace{-1ex}  
     \caption{\small The TCS-SSN performance vs. the road-network distance threshold $\sigma$.}
     \label{fig:sigma}
\end{figure}

\noindent {\bf Measures.} In order to evaluate the performance of our TCS-SSN approach, we report the CPU time and the I/O cost. In particular, the CPU time measures the time cost of  retrieving the TCS-SSN answer candidates by traversing the index (as illustrated in Algorithm \ref{alg:TCS-SSN_processing}), whereas the I/O cost is the number of page accesses during the TCS-SSN query answering.

\noindent {\bf Competitors:} To the best of our knowledge, prior works did not study the problem of community search (CS) over spatial-social networks by considering $(k, d)$-truss communities with user-specified topic keywords, high influences among users, and small road-network distances among users. Thus, we develop three baseline algorithms, $Greedy$, $SIndex$, and $RIndex$.


$Greedy$ first runs the BFS algorithm to retrieve all users with social distance less than $d$ from the query vertex $q$ in social networks. Meanwhile, it prunes those users without any query keywords in $K_q$. Then, it runs another BFS algorithm over road networks to filter out all users with average spatial distance to $q$ greater than $\sigma$. After that, we iteratively apply the pruning on social-network edges for $k$-truss and under other constraints (e.g., influence score, social distance, and spatial distance), and refine/return the resulting connected subgraph.

The $SIndex$ baseline offline constructs a tree index over social-network users and their corresponding social information (e.g., truss values and social-distance information via pivots). In particular, it first partitions users on social networks into subgraphs, which can be treated as leaf nodes, and then recursively groups connected subgraphs in leaf nodes into non-leaf nodes until a final root is obtained. For online TCS-SSN query, $SIndex$ traverses this social-network index by applying the pruning w.r.t. the social-network distance $d$ and the truss value $k$, and refine the resulting subgraphs, similar to the refinement step in Algorithm \ref{alg:TCS-SSN_processing}.

The third baseline, $RIndex$, offline constructs an $R^*$-tree over users' spatial and textual information on road networks. Specifically, we first divide social-network users into partitions based on (1) spatial closeness and (2) keyword information. Then, we treat each partition as a leaf node of the $R^*$-tree, whose spatial locations are enclosed by a minimum bounding rectangles (MBRs). This way, we can build an $R^*$-tree with aggregated keyword information in non-leaf nodes. $RIndex$ traverses the $R^*$-tree and applies pruning based on spatial distance (via pivots) and textual keywords. Finally, the retrieved users (with spatial closeness and keywords) will be refined, as mentioned in the refinement step of Algorithm \ref{alg:TCS-SSN_processing}.

\noindent {\bf Experimental Setup:} Table \ref{table:parameter} depicts the parameter settings in our experiments, where bold numbers are default parameter values. In each set of our subsequent experiments, we will vary one parameter while setting other parameters to their default values. We ran our experiments on a machine with Intel(R) Core(TM) i7-6600U CPU @ 2.60GHz (4 CPUs), ~2.8GHz and 32 GB memory. All algorithms were implemented by C++.

\begin{figure}[t!]\hspace{-4ex}
\subfigure[][{\small CPU time}]{                    
\scalebox{0.3}[0.3]{\includegraphics{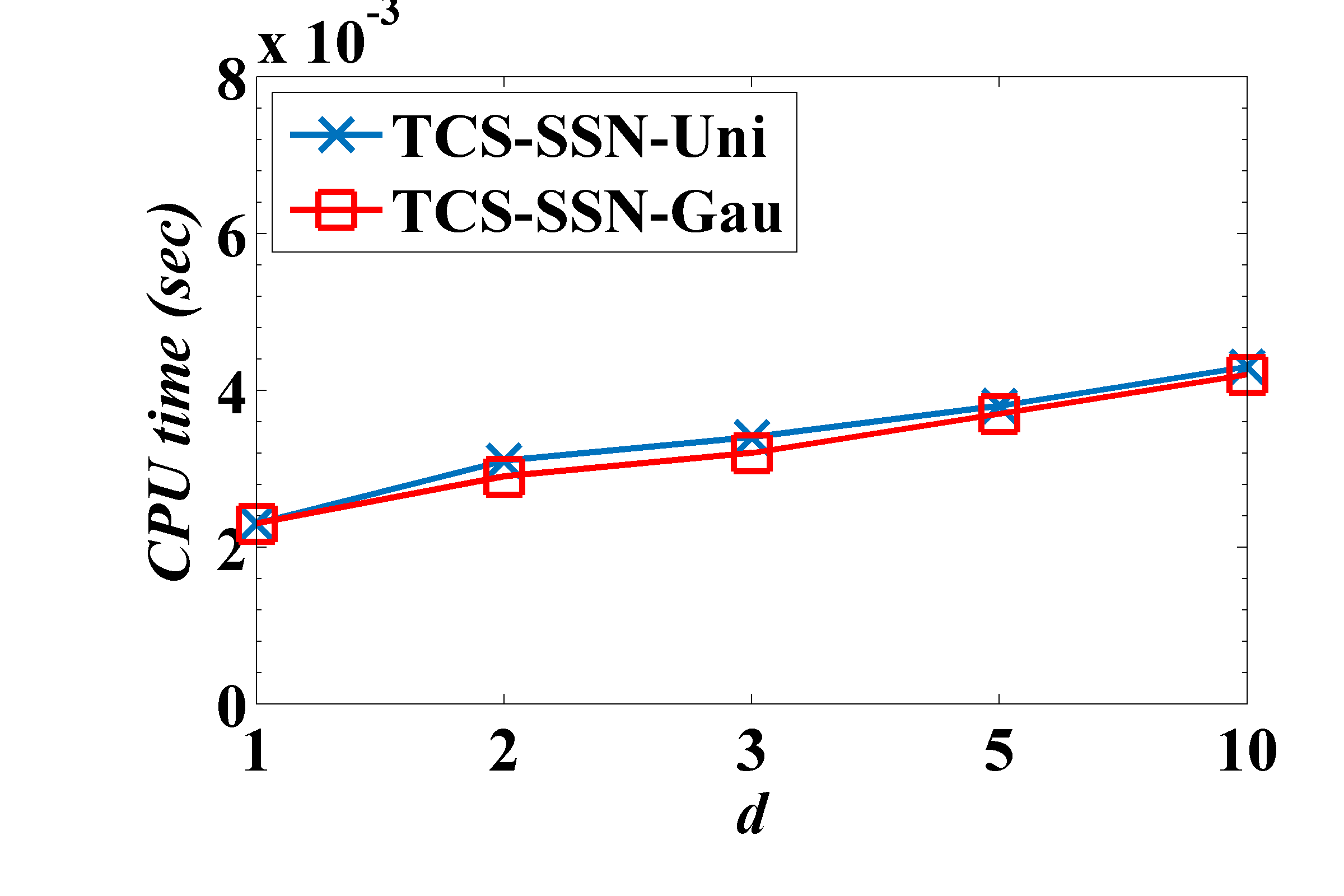}}\label{subfig:d_time}
}%
\subfigure[][{\small I/O cost}]{                    
\scalebox{.3}[.3]{\includegraphics{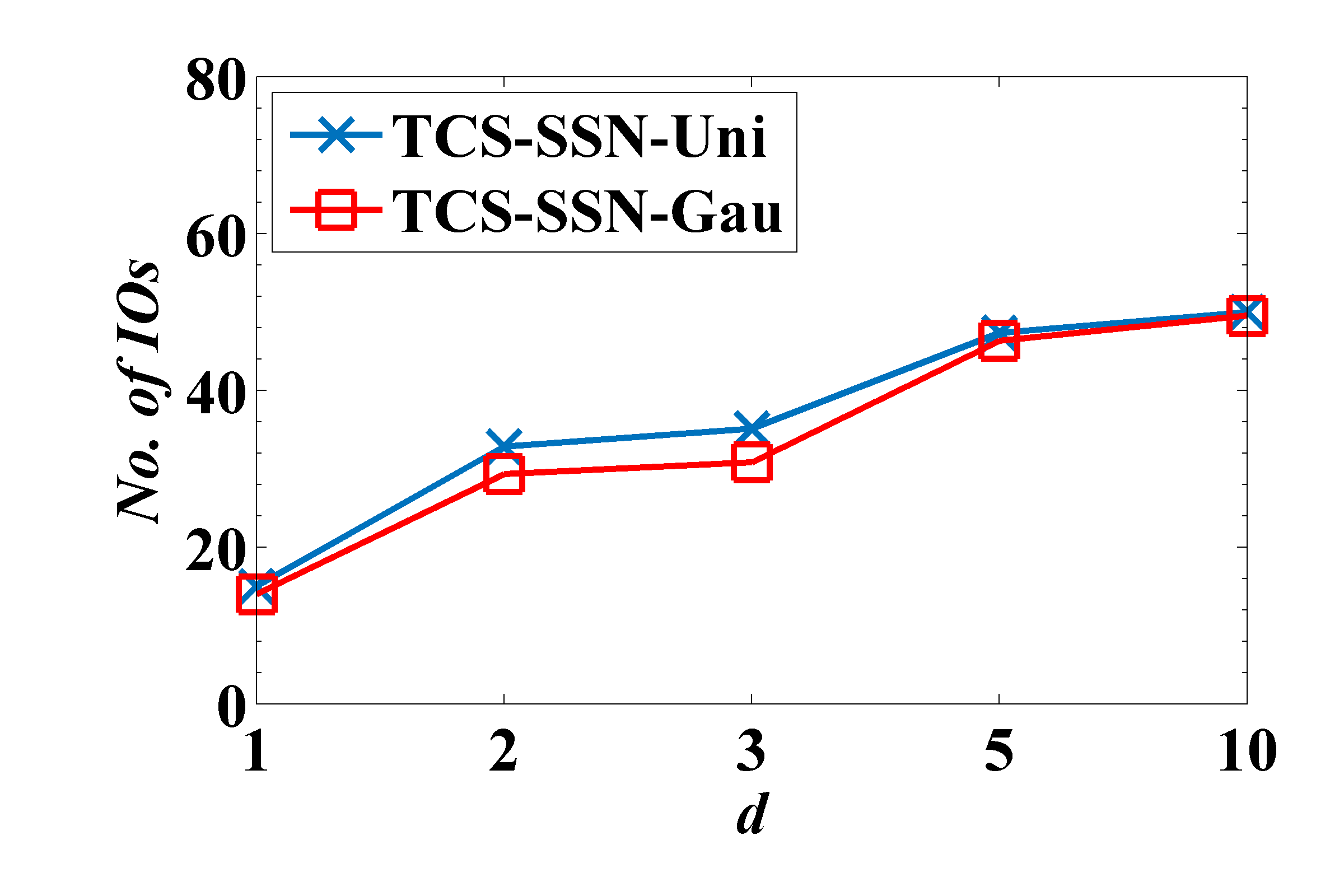}}\label{subfig:d_IO}
}\vspace{-1ex}  
     \caption{\small The TCS-SSN performance vs. the social-network distance (No. of hops) threshold $d$.}
     \label{fig:d}
\end{figure}

\begin{figure}[t!]\hspace{-4ex}
\subfigure[][{\small CPU time}]{                    
\scalebox{0.3}[0.3]{\includegraphics{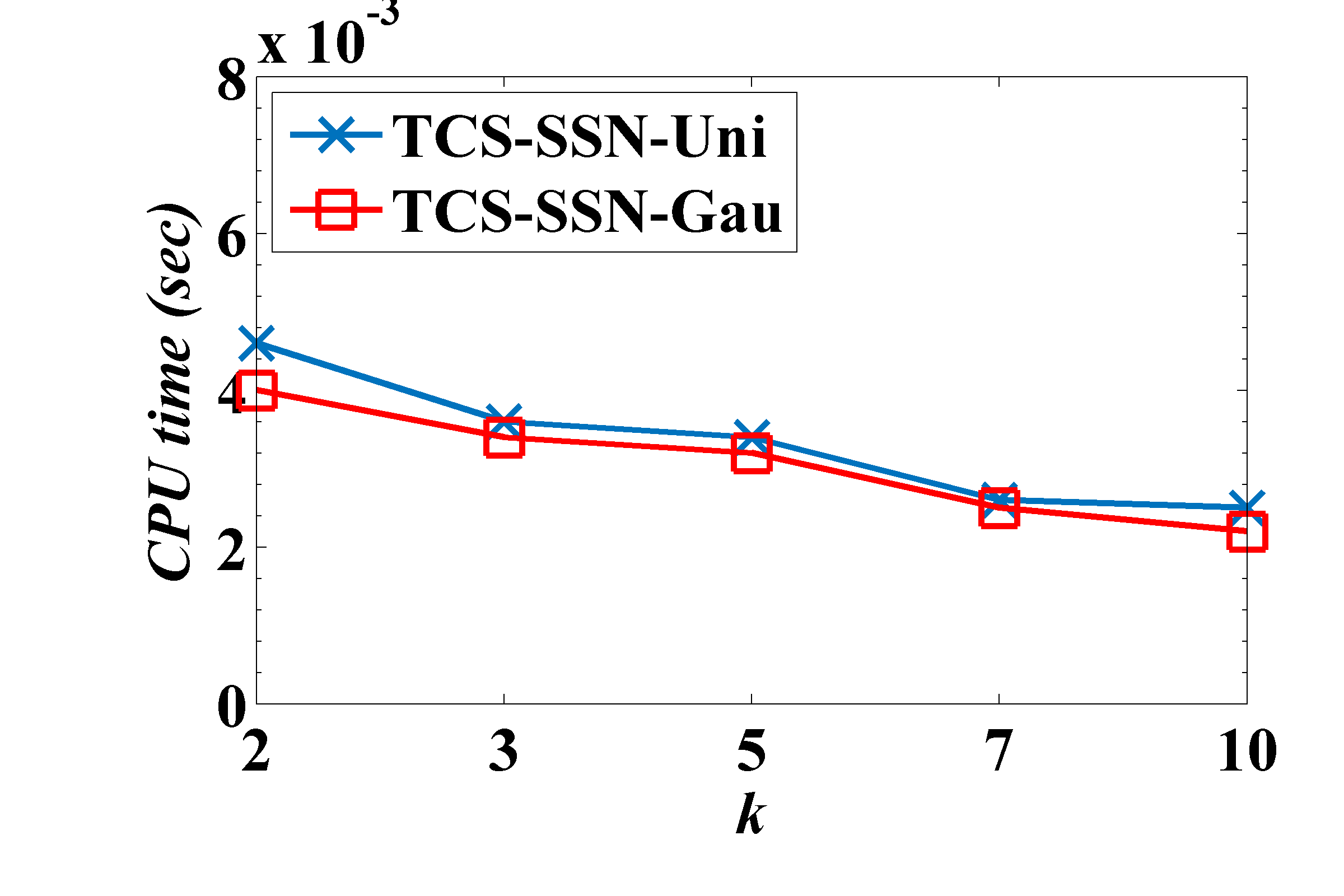}}\label{subfig:k_time}
}%
\subfigure[][{\small I/O cost}]{                    
\scalebox{.3}[.3]{\includegraphics{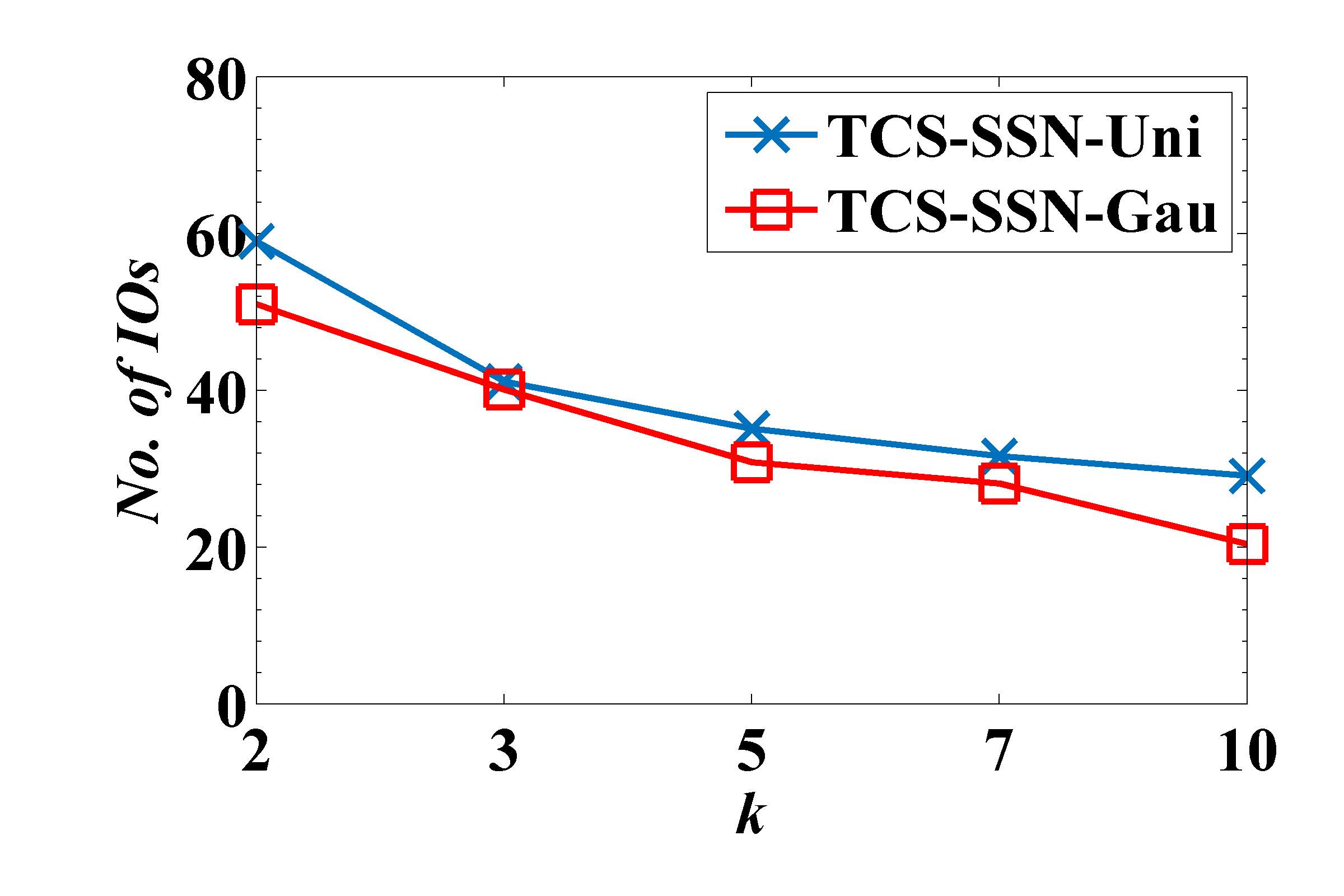}}\label{subfig:k_IO}
}\vspace{-1ex}  
     \caption{\small The TCS-SSN performance vs. the triangle number threshold $k$.}
     \label{fig:k}
\end{figure}

\subsection{TCS-SSN Performance Evaluation}

\noindent {\bf The TSC-SSN Performance vs. Real/Synthetic Data Sets.}
Figure \ref{fig:vs} compares the performance of our TCS-SSN query processing algorithm with three baseline algorithms $Greedy$, $SIndex$, and $RIndex$ over synthetic and real data sets, $Uni$, $Gau$, $Gow\&Cali$, and $Twi\&SF$, in terms of the CPU time and I/O cost, where we set all the parameters to their default values in Table \ref{table:parameter}.
From the experimental results, we can see that our TCS-SSN approach outperforms baselines $Greedy$, $SIndex$, and $RIndex$. This is because TCS-SSN applies effective pruning methods with the help of the social-spatial index.
In particular, for all the real/synthetic data, the CPU time of our proposed TCS-SSN algorithm is $0.0035\sim0.028$ $sec$, and the number of I/Os is around $35\sim162$, which are much smaller than any of the three baseline algorithms $Greedy$, $SIndex$, and $RIndex$. Therefore, this confirms the effectiveness of our proposed pruning strategies and the efficiency of our TCS-SSN query answering algorithm on both real and synthetic data.


Figure \ref{fig:candSize} evaluates the number of the remaining candidate users after the index traversal (applying the pruning methods) over synthetic/real data, where all the parameters are set to their default values. From the figure, we can see that the number of candidate users varies from 5 to 8. This indicates that we can efficiently refine candidate communities with a small number of users.

In Figure \ref{fig:index_time}, we evaluate
the index construction time and space cost of our proposed social-spatial index and the two index-based baselines $SIndex$ and $RIndex$ over $Uni$, $Gau$, $Gow\&Cali$, and $Twi\&SF$ data sets.
Figure \ref{subfig:indexTime} demonstrates the index construction time for our proposed social-spatial index and the baselines $SIndex$ and $RIndex$. 
For $Twi\&SF$ data set (with over than $2.1M$ edges), the index construction time of our social-spatial index takes around 45 minutes. 
The majority of this time cost goes to the computation of the maximum edge support for all edges in the graph, $sup(e)$, which takes $O(E(G_s)^{1.5})$ by applying Wang et al. \cite{wang2012truss}. 
Note that, the social-spatial index (as well as $SIndex$ and $RIndex$ indexes) is ofline constructed only once. 
Furthermore, Figure \ref{subfig:indexUp} shows the index space cost of our proposed social-spatial index and the two baselines $SIndex$ and $RIndex$. From the experimental results, our social-spatial index is much more space efficient than $RIndex$ that uses $R^*$-tree, and is comparable to $SIndex$.


To show the robustness of our TCS-SSN approach, in subsequent experiments, we will vary different parameters (e.g., $\sigma$, $d$, $k$, $\theta$, and so on, as depicted in Table \ref{table:parameter}) on synthetic data sets, $Uni$ and $Gau$.

\noindent {\bf Effect of the Road-Network Distance Threshold $\sigma$.} Figure \ref{fig:sigma} shows the performance of our TCS-SSN approach, by varying the road-network distance threshold $\sigma$ from $0.5$ mile to $5$ miles, where default values are used for other parameters. When $\sigma$ increases, more social-network users will be considered, and thus the CPU time and I/O cost will increase. Nevertheless, for different $\sigma$ values, both CPU time and I/O cost remain low ($0.002 \sim 0.0075$ $sec$ and $17 \sim 64$ I/Os, respectively).


\noindent {\bf Effect of the Social-Network Distance Threshold $d$.} 
Figure \ref{fig:d} varies the social-distance threshold $d$ (i.e., the threshold for the number of hops) from 1 to 10, and reports the CPU time and I/O cost of our TCS-SSN approach over $Uni$ and $Gau$ data sets, where other parameters are set to their default values. With the increase of the social-distance threshold $d$, more candidate communities (with more social-network users) will be retrieved for evaluation. Therefore, the CPU time and I/O cost become higher for larger threshold $d$. Nonetheless, for different $d$ values, the CPU time remains small (i.e., around $0.0023\sim 0.005$ $sec$), and the I/O cost is low (with $16 \sim 59$ page accesses).

\noindent {\bf Effect of the Triangle Number Threshold $k$.} 
Figure \ref{fig:k} examines the TCS-SSN performance with different thresholds $k$ for the number of triangles, in terms of the CPU time and I/O cost, where $k = 2, 3, 5, 7,$ and $10$, and other parameters are set to their default values. In figures, when $k$ becomes large, many users with low degrees (i.e., $< k$) will be safely pruned, and thus the CPU time and I/O cost are expected to reduce substantially (as confirmed by figures).  Nevertheless, the CPU time and the I/O cost remain low (i.e., about $0.002\sim0.0065$ $sec$ and $20\sim 61$ I/Os, respectively), which indicates the efficiency of our proposed TCS-SSN approach for different $k$ values.


\begin{figure}[t!]\hspace{-4ex}
\subfigure[][{\small CPU time}]{                    
\scalebox{0.3}[0.3]{\includegraphics{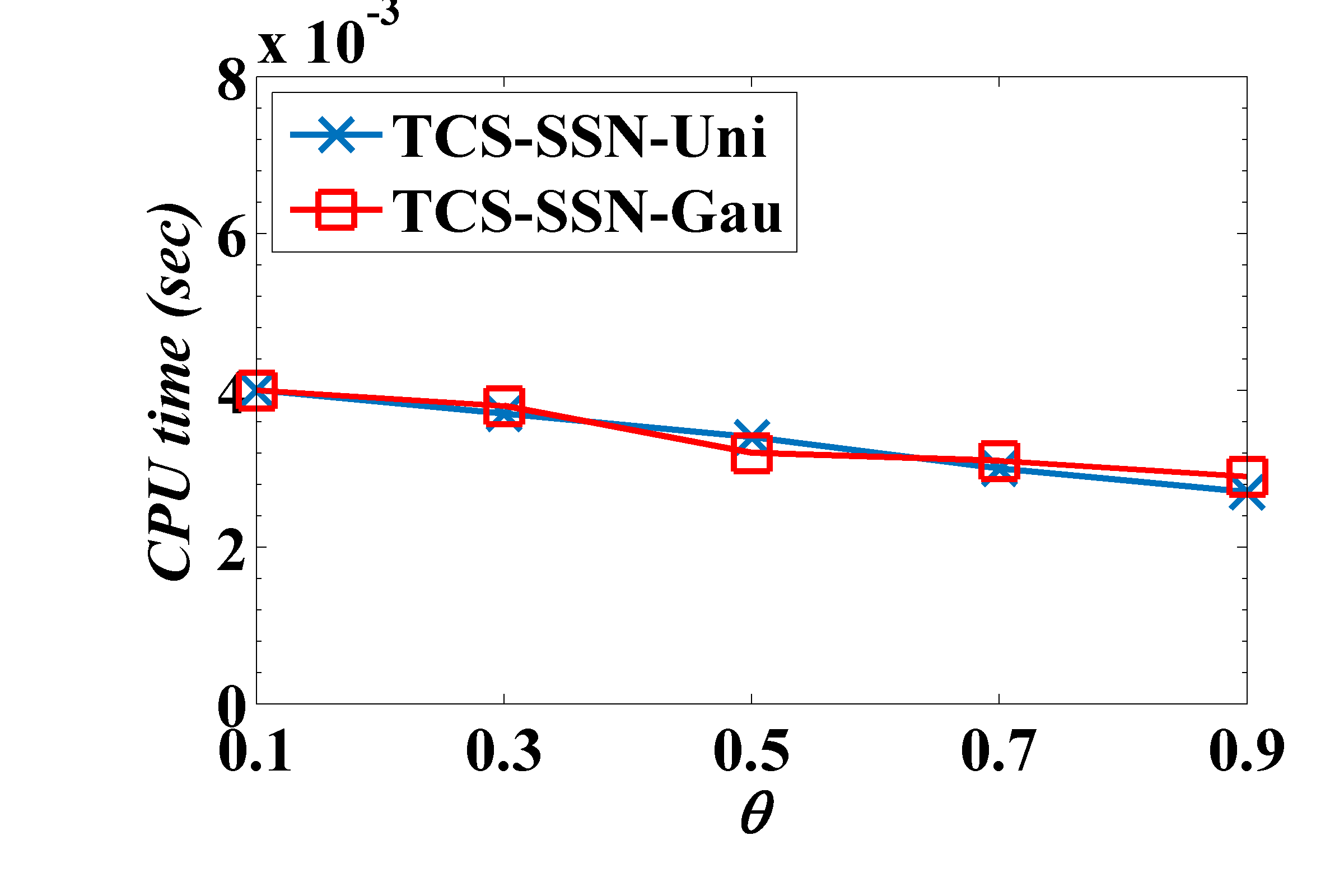}}\label{subfig:theta_time}
}%
\subfigure[][{\small I/O cost}]{                    
\scalebox{.3}[.3]{\includegraphics{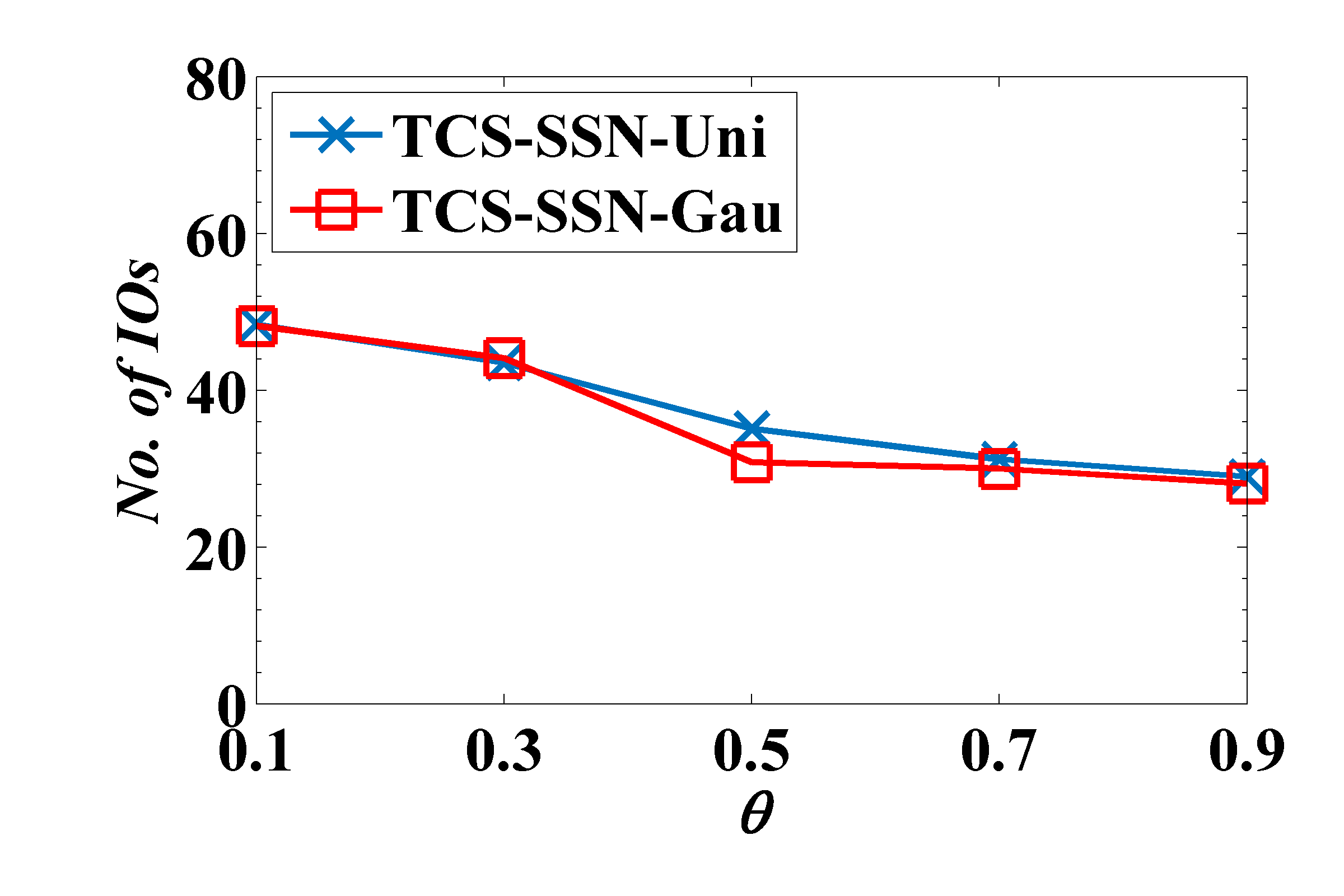}}\label{subfig:theta_IO}
}\vspace{-1ex}  
     \caption{\small The TCS-SSN performance vs. influence score threshold $\theta$.}
     \label{fig:theta}
\end{figure}

\begin{figure}[t!]\hspace{-4ex}
\subfigure[][{\small CPU time}]{                    
\scalebox{0.3}[0.3]{\includegraphics{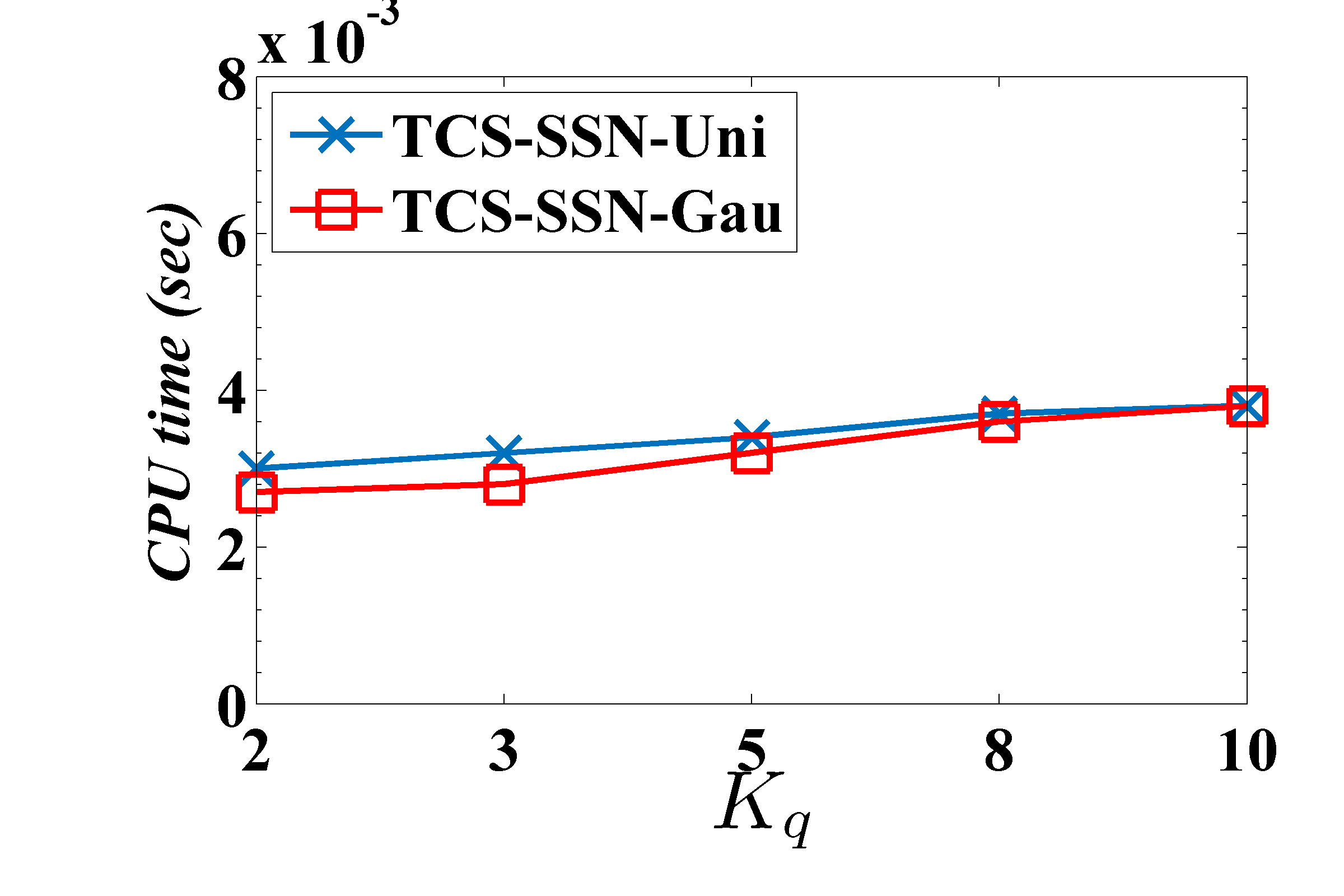}}\label{subfig:Kq_time}
}%
\subfigure[][{\small I/O cost}]{                    
\scalebox{.3}[.3]{\includegraphics{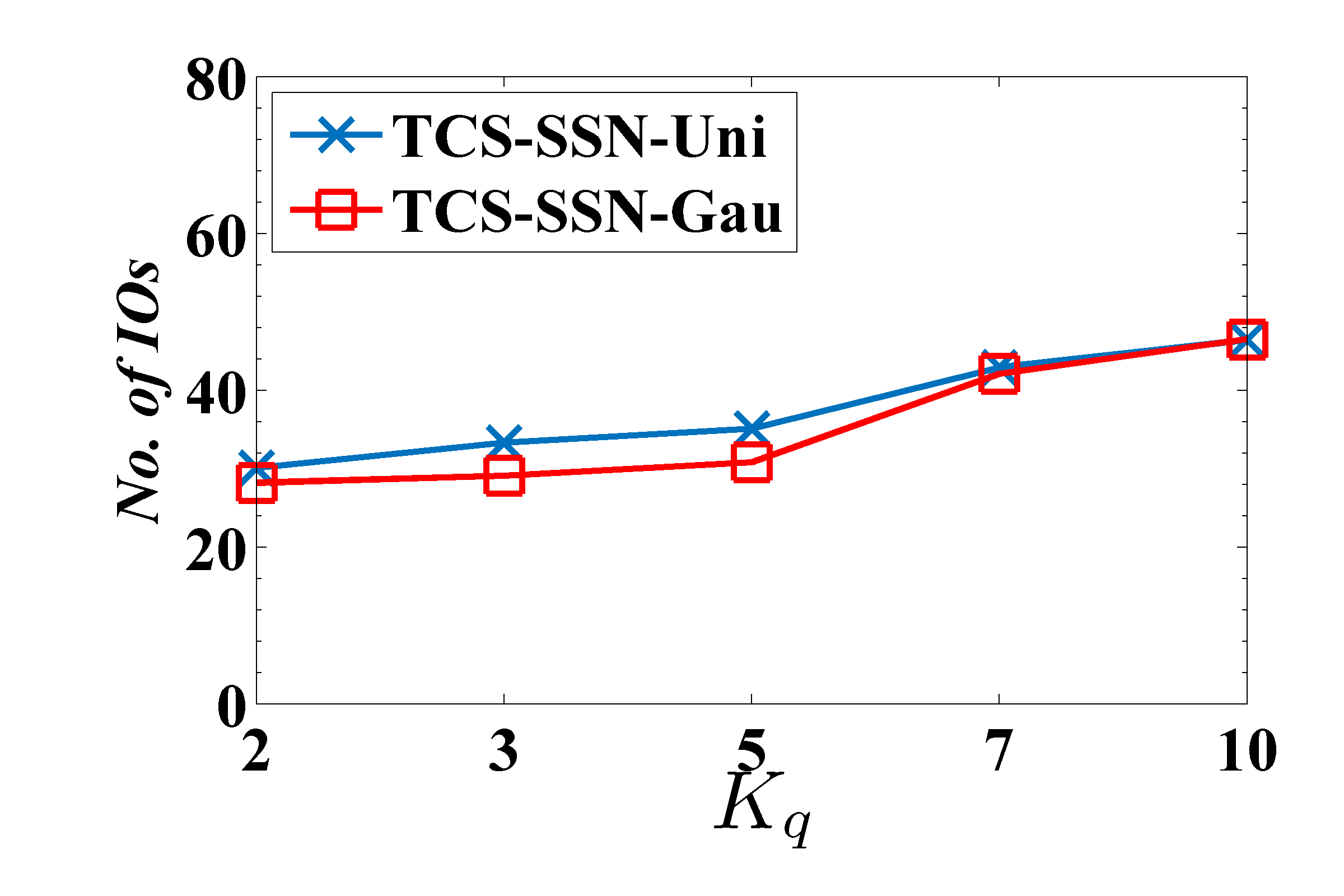}}\label{subfig:Kq_IO}
}\vspace{-1ex}  
     \caption{\small The TCS-SSN performance vs. the size, $|K_q|$, of the keyword query set.}
     \label{fig:Kq}
\end{figure}

\noindent {\bf Effect of the Influence Score Threshold $\theta$.} Figure \ref{fig:theta} illustrates the CPU time and the I/O cost of our TCS-SSN approach by varying the interest score threshold $\theta$ from 0.1 to 0.9, where all other parameter values are set by default. From the experimental results, we can see that both CPU time and I/O cost smoothly decrease with large $\theta$ values. This is because larger $\theta$ can filter out more edges with low influence scores, which leads to less user candidates for the filtering and refinement. Nonetheless, the time and I/O cost of our TCS-SSN approach remain low (i.e., $0.0025\sim0.0041$ $sec$ for the CPU time and $25\sim50$ page accesses).

\noindent {\bf Effect of the Size, $|K_q|$, of the Keyword Query Set.}
Figure $\ref{fig:Kq}$ demonstrates the performance of our TCS-SSN approach with different numbers of query keywords in $K_q$, where $|K_q|= 2, 3, 5, 8$, and $10$, and default values are used for other parameters. Intuitively, when $|K_q|$ becomes larger (i.e., more query keywords), we need to consider more potential users, which incurs higher CPU time and I/O cost. Despite that, the CPU time and I/O cost of our TCS-SSN approach remain low (i.e., $0.0025\sim0.004$ $sec$ for the CPU time and $25\sim49$ page accesses).

\begin{figure}[t!]\hspace{-4ex}
\subfigure[][{\small CPU time}]{                    
\scalebox{0.3}[0.3]{\includegraphics{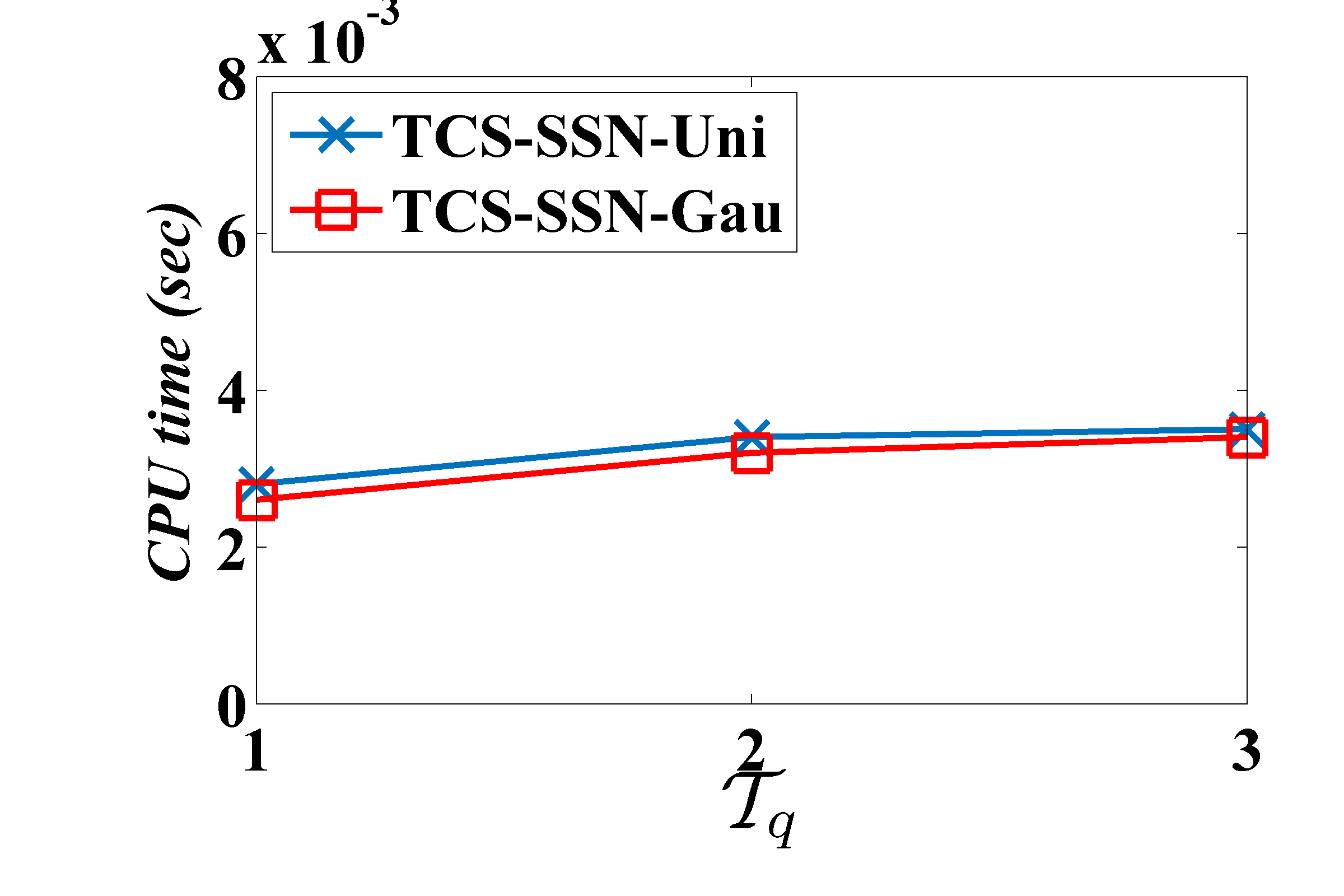}}\label{subfig:Tq_time}
}%
\subfigure[][{\small I/O cost}]{                    
\scalebox{.3}[.3]{\includegraphics{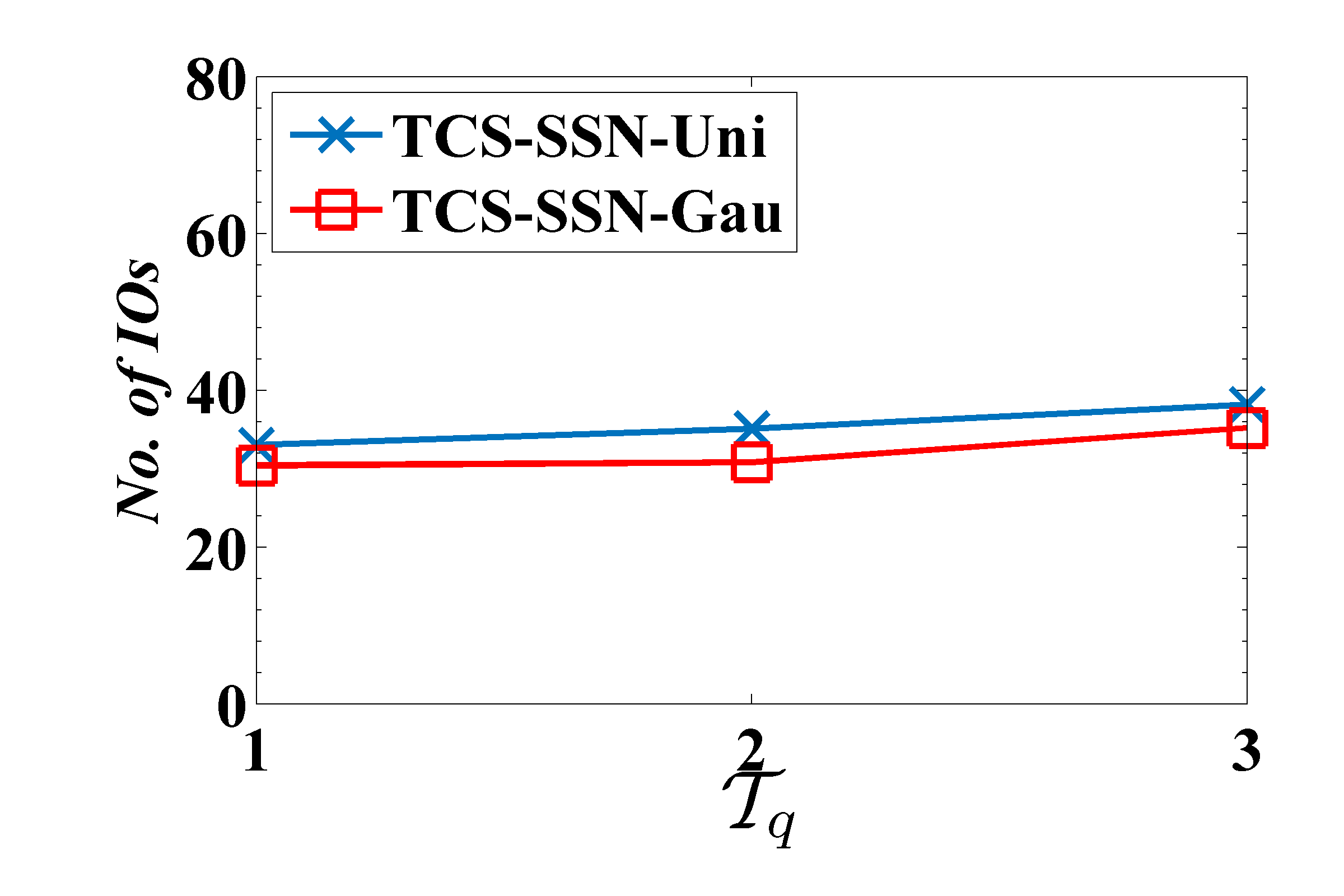}}\label{subfig:Tq_IO}
}\vspace{-1ex}  
     \caption{\small The TCS-SSN performance vs. the size, $|\mathcal{T}_q|$, of the topic query set.}
     \label{fig:Tq}
\end{figure}

\begin{figure}[t!]\hspace{-4ex}
\subfigure[][{\small CPU time}]{                    
\scalebox{0.26}[0.3]{\includegraphics{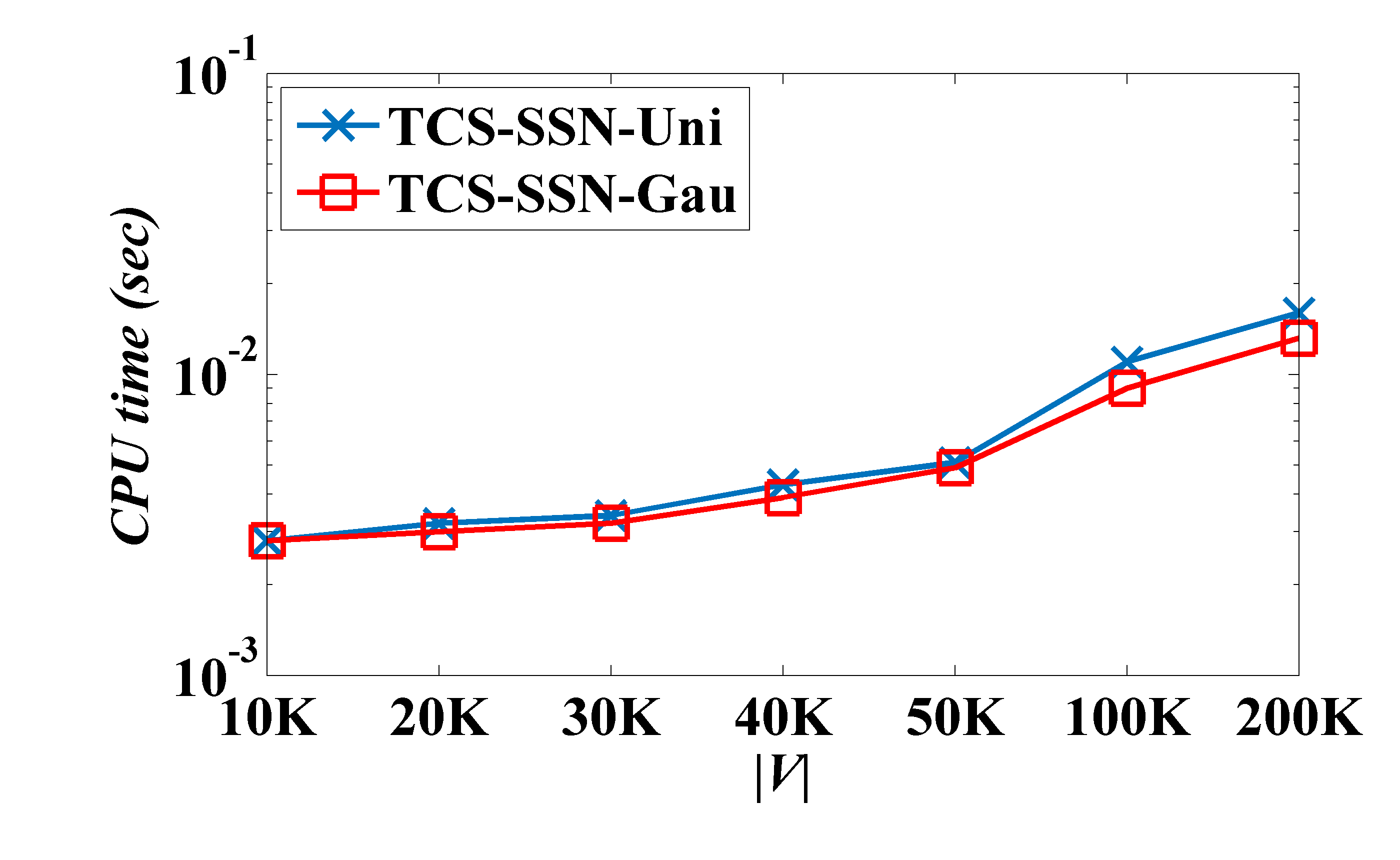}}\label{subfig:V_time}
}%
\subfigure[][{\small I/O cost}]{                    
\scalebox{.26}[.3]{\includegraphics{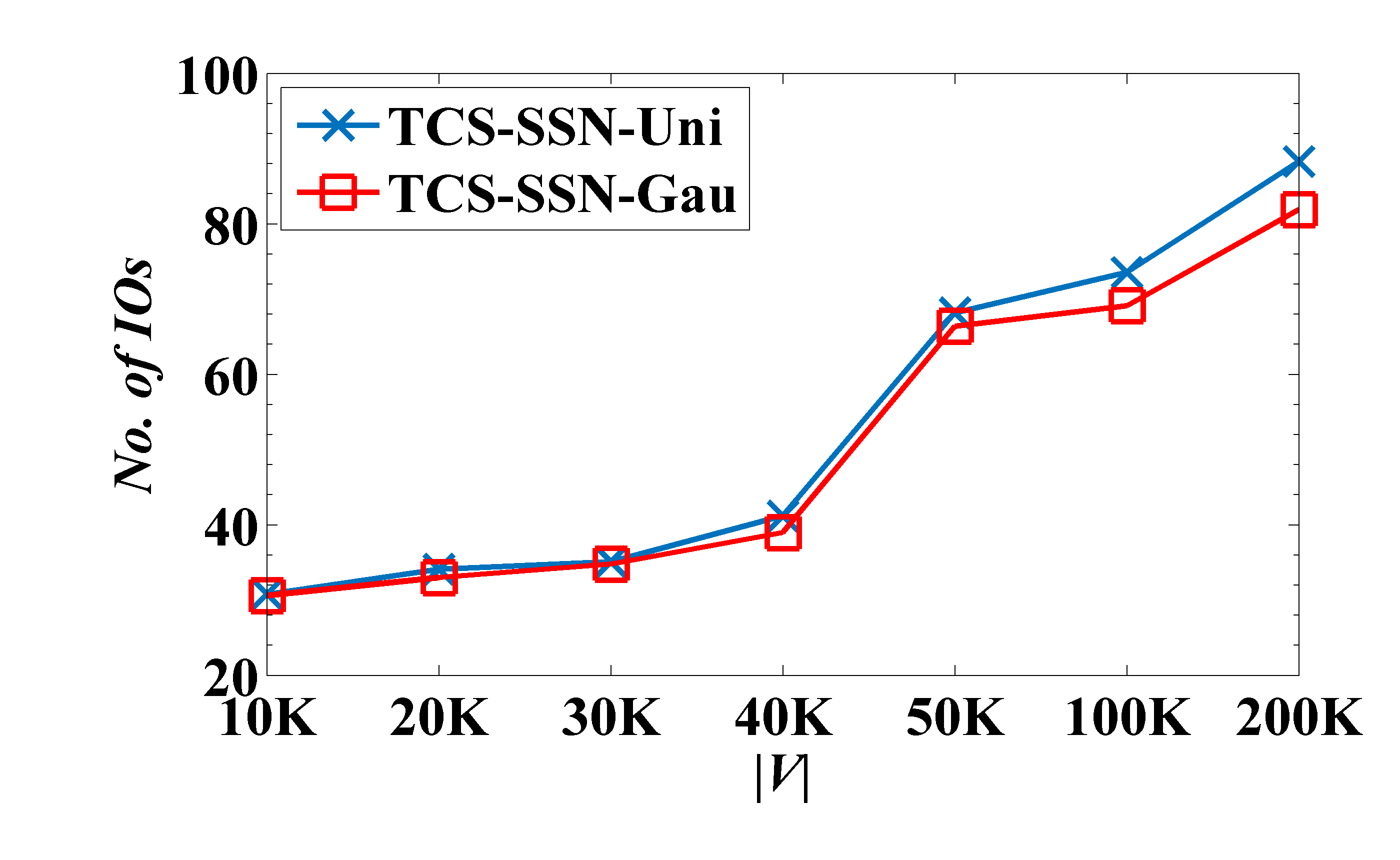}}\label{subfig:V_IO}
}\vspace{-1ex}  
     \caption{\small The TCS-SSN performance vs. the number, $|V(G_r)|$ (or $|V(G_s)|$), of vertices in spatial or social networks.}
     \label{fig:V}
\end{figure}

\noindent {\bf Effect of the Size, $|\mathcal{T}_q|$, of the Topic Query Set.}
Figure \ref{fig:Tq} illustrates the performance of our TCS-SSN approach by varying the number of query topics (in $\mathcal{T}_q$) on edges, where $|\mathcal{T}_q|= 1, 2$, and $3$, and other parameters are set to their default values.
The experimental results show that the TCS-SSN performance is not very sensitive to $|\mathcal{T}_q|$.
The CPU time remains low (i.e., $0.003\sim0.0035$ $sec$) and the I/O cost is around $30 \sim 38$, which indicate the efficiency of our TCS-SSN approach with different $|\mathcal{T}_q|$ values.

\noindent {\bf Effect of the Number, $|V(G_r)|$ (or $V(G_s)$), of Vertices in Road (Social) Networks.}
Figure \ref{fig:V} shows the scalability of our TCS-SSN approach with different sizes of spatial/road networks, $|V(G_r)|$ (or $|V(G_s)|$), of spatial/road networks (denoted as $|V|$), where $|V|$ varies from $10K$ to $200K$, and other parameters are set to their default values. In figures, when the number of road-network (or social-network) vertices increases, both CPU time and I/O cost smoothly increase. Nevertheless, the CPU time and I/O costs of our TCS-SSN approach remain low (i.e., $0.0028\sim0.017$ $sec$ for the time cost and $30\sim 89$ page accesses, respectively), which confirms the scalability of our TCS-SSN approach against large network sizes.

\begin{figure}[t!]\hspace{-4ex}
\subfigure[][{\small}]{                    
\scalebox{0.35}[0.35]{\includegraphics{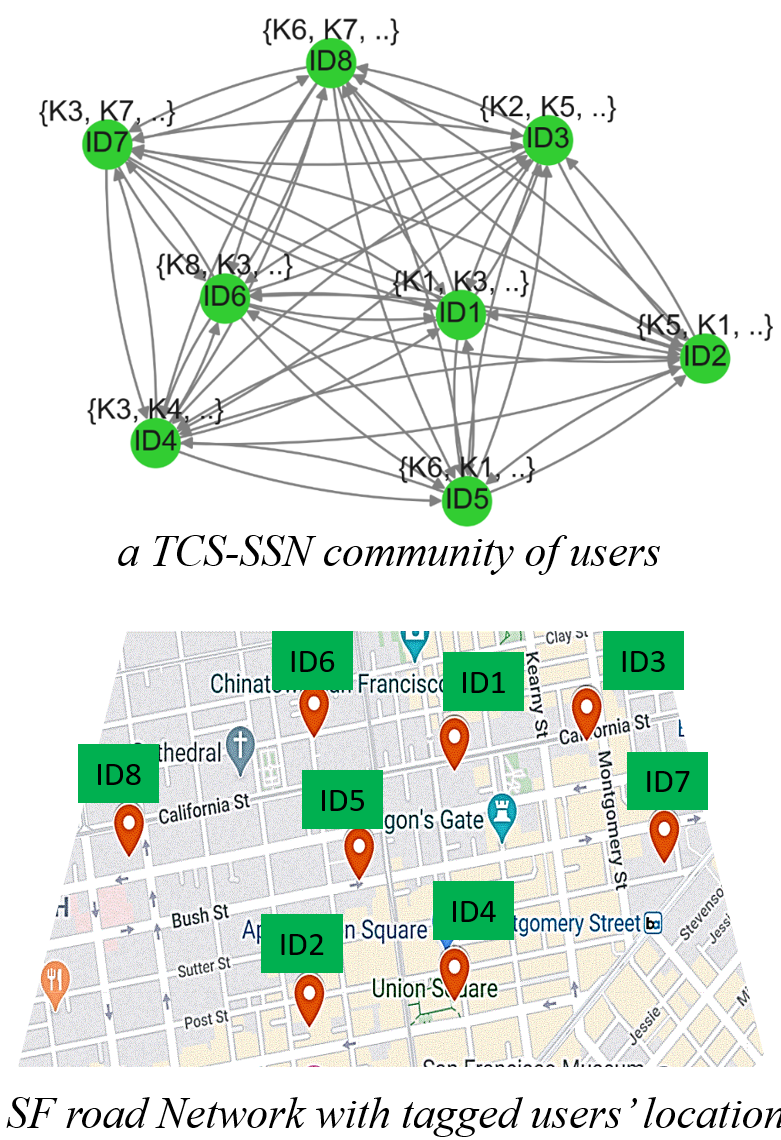}}\label{subfig:study}
}%
\subfigure[][{\small }]{                    
\scalebox{0.32}[0.32]{\includegraphics{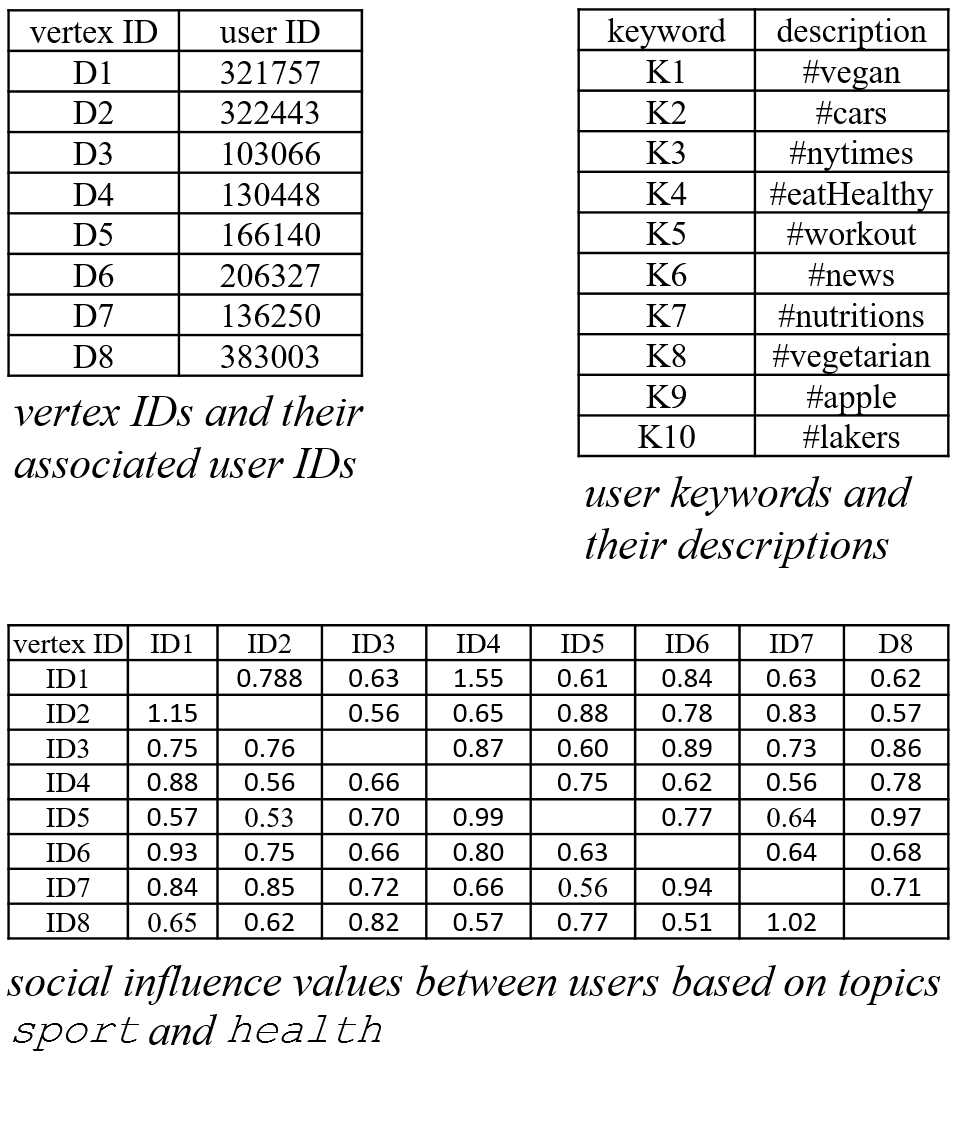}}\label{subfig:tbl}
}\vspace{-1ex}  
     \caption{\small A TCS-SSN case study on spatial-social networks $Twi\&SF$.}
     \label{fig:CaseStudy}
\end{figure}

\subsection{A Case Study}
Finally, we conduct a case study of our TCS-SSN problem on real-world spatial-social networks, $Twi\&SF$ (i.e., Twitter \cite{li2014efficient} with San Francisco road networks \cite{li2005trip}). As illustrated in Figure \ref{subfig:study}, each social-network user is associated with keywords (i.e., Twitter hashtags) such as $K1 \sim K10$ from user accounts, and has checkin locations on road networks, where the user IDs of vertices and descriptions of keywords are depicted in Figure \ref{subfig:tbl}. 

Assume that we have a TCS-SSN query over $Twi\&SF$, where $ID2$ is a query vertex, truss value $k=5$, social-network distance threshold $d = 3$, spatial-distance threshold $\sigma = 2$ $miles$, influence score threshold $\theta = 0.5$, query topic set $\mathcal{T}_q = $ (\texttt{sport}, \texttt{health}), and query keyword set $K_q = \{$\textit{vegan}, \textit{vegetarian}, \textit{eatHealthy}, \textit{workout}, \textit{nutritions}$\}$. Figure \ref{subfig:study} shows the resulting TCS-SSN community, which contains a subgraph of 8 users, $ID1\sim ID8$. Each user in this community is associated with at least one query keyword in $K_q$, and they show strong structural connectivity (satisfying the $(5, 3)$-truss constraints) and spatial closeness on road networks ($\leq 2$ $miles$). Moreover, Figure \ref{subfig:tbl} depicts an influence matrix (w.r.t. $\mathcal{T}_q$) for pairwise users in this community, each element of which is above influence threshold $\theta$ (i.e., $0.5$). This confirms their high influences to each other within our retrieved TCS-SSN community.


\section{Related Work}
\label{sec:related_work}
\noindent {\bf Community Detection.} The community detection problem aims to discover all communities in a large-scale graph such as social networks or bibliographic networks. Some prior works \cite{fortunato2010community, newman2004finding} retrieved communities in large graphs, by considering link information only. More recent work was carried by \cite{xu2012model, liu2009topic,nallapati2008joint, zhou2009graph}, that devoted for attribute graphs, by using clustering techniques. In \cite{zhou2009graph}, for example, the links and keyword vertices are considered to compute the pairwise vertex similarity in order to cluster the large graph. Zang et al. \cite{chopade2016framework} proposed a framework that applies a game-theoretic approach to identify dense communities in large-scale complex networks. Recently, other works carried by \cite{girvan2002community, expert2011uncovering, guo2008regionalization, chen2015finding} focused on detecting communities in spatially constrained graphs, where graph vertices are associated with spatial coordinates. 

In the aforementioned works, a geo-community is defined as a community, in which vertices are densely connected and loosely connected with other vertices. The resulting communities are more compact in geographical space. Techniques such as {\it average linkage measure} \cite{guo2008regionalization} and {\it modularity maximization} \cite{expert2011uncovering, chen2015finding} are applied to discover geo-communities. However, Lancichinetti et al. \cite{lancichinetti2011limits} argued that modularity-based methods often fail to resolve small-size communities. 

\noindent {\bf Community Search.} Community search problem (CS) aims to obtain communities in an ``online'' manner, based on a query request. Several existing works \cite{sozio2010community, cui2014local, cui2013online, li2015influential,huang2015approximate} have proposed efficient algorithms to obtain a community starting from and including a query vertex $q$. In \cite{sozio2010community, cui2014local}, the {\it minimum degree} is used to measure the structure cohesiveness of community. Sozio et al. \cite{sozio2010community} proposed the first algorithm $Global$ to obtain $k$-core community containing a query vertex $q$. Cui et al. \cite{cui2014local} used local expansion techniques to boost the query performance.
Furthermore, Li et al. \cite{li2017most} proposed the most influential community search over large social networks to disclose the $\mathcal{C}^{kr}$ community with the highest outer influences, where the $\mathcal{C}^{kr}$ community contains at least $k$ nodes, and any two nodes in $\mathcal{C}^{kr}$
can be reached at most $r$ hops. 
Bi et al. \cite{bi2018optimal} proposed an optimal approach to retrieve top-$k$ influential communities (subgraphs), such that each subgraph $g$ is a maximal connected subgraph with minimum degree of at least $\gamma$, and has the highest influence value.
Akbas et al. \cite{akbas2017truss} introduced a truss-based indexing approach, where they can in optimal time detect the $k$-truss communities in large network graphs.
Fang et al. \cite{fang2020effective} studied the community search problem over large heterogeneous information networks, that is, given a query vertex
$q$, find a community from a heterogeneous network containing $q$, such that all the vertices are with the same type of $q$ and have close relationships, where the relationship between two vertices of the same type is modeled by a $meta-path$, and the cohesiveness of the community is measured by the classic minimum degree metric with the $meta-path$.
Note that, the aforementioned previous works \cite{li2017most, bi2018optimal, akbas2017truss, fang2020effective} did not consider spatial cohesiveness, topic-related social influences, nor keywords, which different from our proposed TCS-SSN problem.

Some other works \cite {fang2016effective, li2015influential} used minimum degree metric to search communities for attribute graphs. Other well-known structure cohesiveness $k$-clique \cite{cui2013online}, $k$-truss \cite{huang2015approximate} have also been considered for online community search. However, these works are designed for non-spatial graphs. Huang et al. \cite{huang2017attribute} proposed $(k, d)$-truss for geo-spatial networks, but they did not take into account user topic keywords, social influence, neither road-network distance.

\noindent {\bf Geo-Social Networks.} Query processing on location-based social networks has become increasingly important in many real applications. Yang et al. \cite{yang2012socio} studied the problem of \textit{socio-spatial group query} (SSGQ), which retrieves a group of connected users (friends) with the smallest summed distance to a given query point $q$. Li et al. \cite{Li12} proposed another query type that retrieves a group of $k$ users who are interested in some given query keywords and are spatially close to each other. Yuan et al. \cite{Yuan16} studied the $k$NN query which obtains $k$ POIs that are not only closest to query point $q$, but also recommended by one's friends on social networks under the IC model.
Fang et al. \cite{fang2017effective} introduced the spatial-aware community (or SAC), which retrieves a subgraph (containing a given query vertex $q$) from geo-social networks that has high structural cohesiveness and spatial cohesiveness. \cite{GPSSN} proposed the GP-SSN query that retrieves a set $S$ of users from social networks and a set $R$ of potential POIs from road networks to be visited by the set of users in $S$. 
Chen et al. \cite{chen2018maximum} discussed the co-located community search, that is a subgraph satisfying connectivity, structural cohesiveness, and spatial cohesiveness.
Chen et al. \cite{chen2018maximum} considered communities that match query predicates and have the maximum cardinality globally, whereas Fang et al. \cite{fang2017effective} focused on finding a locally optimal community containing a query vertex. 
Previous works on geo-social community search neglected the social influence among users and road-network distance. In our proposed TCS-SSN problem, we introduce a new and different definition of communities that not only are spatially and socially close, but also have high social influence and small driving distance among community members.

\noindent {\bf Keyword Search and Spatial Keyword Queries.}
The keyword search problem has been extensively studied in both domains of relational databases and graphs. Given a set of query keywords, the keyword search in relational databases \cite{chaudhuri2004system, kacholia2005bidirectional,hristidis2003efficient}
usually finds a minimal connected tuple tree that contains all the query keywords.
In graph databases \cite{li2008ease, qin2009querying}, the keyword search problem retrieves a subgraph containing the given query keywords.
Furthermore, in spatial databases containing both spatial and textual information, another interesting problem is the spatial keyword query, which returns relevant POIs that both satisfy the spatial query predicates and match the given query keywords.
Coa et al.~\cite{cao2012spatial} categorized the spatial keyword queries based on their ways of specifying spatial and textual predicates, including Boolean Range Queries \cite{hariharan2007processing}, Boolean kNN Queries \cite{cary2010efficient, de2008keyword}, and Top-$k$ kNN Queries \cite{cao2010retrieving, cong2009efficient}.
Recently, Zhang et al. \cite{zhang2019keyword} proposed the keyword-centric community search (KCCS) over an attributed graph, which finds a community (subgraph) with the degree of each node at least $k$, and the distance between nodes and all the query keywords being minimized. However, Zhang et al. \cite{zhang2019keyword} did not consider the spatial cohesiveness neither the social influence.
Moreover, Islam et al. \cite{islam2019keyword} proposed the keyword-aware influential community query (KICQ) over an attributed graph, which returns $r$ most influential communities in the attributed graph, such that the returned community has high influence (containing highly influential members) based on certain keywords. An application of KICQ is to find the most influential community of users who are working in “ML” or “DB.” Different from KICQ, our proposed TCS-SSN finds a community that not only has a high social cohesiveness and covers certain keywords, but also has a high spatial cohesiveness. Furthermore, our TCS-SSN problem returns the community with high influence score among community members (with respect to specific topics), rather than members with high influences to others in KICQ
Chen et al. \cite{chen2019contextual} introduced a parameter-free contextual community model for attributed community search. Given an attributed graph and a set of query keywords describing the desired matching community context, their proposed query returns a community that has both structure and attribute cohesiveness w.r.t. the provided query context.
In contrast, different from \cite{chen2019contextual}, our TCS-SSN problem returns the community over spatial-social networks (instead of social networks only), which has high social cohesiveness, spatial cohesiveness, social influence, and covers a set of keywords (rather than measuring the context closeness of the community with the query context). 
Thus, with different underlying data models (relational or graph data) and query types, we cannot directly borrow previous techniques for (spatial) keyword search or community search on attributed graphs to solve our TCS-SSN problem.

To our best knowledge, the TCS-SSN problem has not been studied by prior works on spatial-social networks, which considers $(k, d)$-truss communities with user-specified topic keywords, high influences among users, and small road-network distances among users. Due to different data models and query types, previous techniques on location-based social networks cannot be directly used for tackling our TCS-SSN problem.

\section{Conclusions}
\label{sec:conclusion}

In this paper, we formalize and tackle an important problem, \textit{topic-based community search over spatial-social networks} (TCS-SSN), which retrieves communities of users (including a given query user) that are spatially and socially close to each other. In order to efficiently tackle this problem, we design effective pruning mechanisms to reduce the TCS-SSN problem space, propose a novel social-spatial index over spatial-social networks, and develop efficient algorithms to process TCS-SSN queries. Through extensive experiments, we evaluate the efficiency and effectiveness of our proposed TCS-SSN processing approaches over both real and synthetic data.

\balance
\bibliographystyle{abbrv}
\let\xxx=\bibitem\def\bibitem{\par\vspace{0mm}\xxx}
\bibliography{all}

\begin{thebibliography}{10}

\bibitem{akbas2017truss}
E.~Akbas and P.~Zhao.
\newblock Truss-based community search: a truss-equivalence based indexing
  approach.
\newblock {\em Proceedings of the VLDB Endowment}, 10(11):1298--1309, 2017.

\bibitem{al2019efficient}
A.~Al-Baghdadi, X.~Lian, and E.~Cheng.
\newblock Efficient path routing over road networks in the presence of ad-hoc
  obstacles (technical report).
\newblock {\em arXiv preprint arXiv:1910.04786}, 2019.

\bibitem{GPSSN}
A.~{Al-Baghdadi}, G.~{Sharma}, and X.~{Lian}.
\newblock Efficient processing of group planning queries over spatial-social
  networks.
\newblock {\em IEEE Transactions on Knowledge and Data Engineering}, pages
  1--1, 2020.

\bibitem{armenatzoglou2013general}
N.~Armenatzoglou, S.~Papadopoulos, and D.~Papadias.
\newblock A general framework for geo-social query processing.
\newblock {\em Proceedings of the VLDB Endowment}, 6(10):913--924, 2013.

\bibitem{barbieri2013topic}
N.~Barbieri, F.~Bonchi, and G.~Manco.
\newblock Topic-aware social influence propagation models.
\newblock {\em Knowledge and information systems}, 37(3):555--584, 2013.

\bibitem{bhattacharyya2011analysis}
P.~Bhattacharyya, A.~Garg, and S.~F. Wu.
\newblock Analysis of user keyword similarity in online social networks.
\newblock {\em Social network analysis and mining}, 1(3):143--158, 2011.

\bibitem{bi2018optimal}
F.~Bi, L.~Chang, X.~Lin, and W.~Zhang.
\newblock An optimal and progressive approach to online search of top-k
  influential communities.
\newblock {\em Proceedings of the VLDB Endowment}, 11(9):1056--1068, 2018.

\bibitem{cao2012spatial}
X.~Cao, L.~Chen, G.~Cong, C.~S. Jensen, Q.~Qu, A.~Skovsgaard, D.~Wu, and M.~L.
  Yiu.
\newblock Spatial keyword querying.
\newblock In {\em International Conference on Conceptual Modeling}, pages
  16--29. Springer, 2012.

\bibitem{cao2010retrieving}
X.~Cao, G.~Cong, and C.~S. Jensen.
\newblock Retrieving top-k prestige-based relevant spatial web objects.
\newblock {\em Proceedings of the VLDB Endowment}, 3(1-2):373--384, 2010.

\bibitem{cary2010efficient}
A.~Cary, O.~Wolfson, and N.~Rishe.
\newblock Efficient and scalable method for processing top-k spatial boolean
  queries.
\newblock In {\em International Conference on Scientific and Statistical
  Database Management}, pages 87--95. Springer, 2010.

\bibitem{chaudhuri2004system}
S.~Chaudhuri, S.~Agrawal, and G.~Das.
\newblock System for keyword based searching over relational databases, Oct.~5
  2004.
\newblock US Patent 6,801,904.

\bibitem{chen2019contextual}
L.~Chen, C.~Liu, K.~Liao, J.~Li, and R.~Zhou.
\newblock Contextual community search over large social networks.
\newblock In {\em 2019 IEEE 35th International Conference on Data Engineering
  (ICDE)}, pages 88--99. IEEE, 2019.

\bibitem{chen2018maximum}
L.~Chen, C.~Liu, R.~Zhou, J.~Li, X.~Yang, and B.~Wang.
\newblock Maximum co-located community search in large scale social networks.
\newblock {\em Proceedings of the VLDB Endowment}, 11(10):1233--1246, 2018.

\bibitem{chen2015online}
S.~Chen, J.~Fan, G.~Li, J.~Feng, K.-l. Tan, and J.~Tang.
\newblock Online topic-aware influence maximization.
\newblock {\em Proceedings of the VLDB Endowment}, 8(6):666--677, 2015.

\bibitem{chen2015finding}
Y.~Chen, J.~Xu, and M.~Xu.
\newblock Finding community structure in spatially constrained complex
  networks.
\newblock {\em International Journal of Geographical Information Science},
  29(6):889--911, 2015.

\bibitem{chopade2016framework}
P.~Chopade and J.~Zhan.
\newblock A framework for community detection in large networks using
  game-theoretic modeling.
\newblock {\em IEEE Transactions on Big Data}, 3(3):276--288, 2016.

\bibitem{cong2009efficient}
G.~Cong, C.~S. Jensen, and D.~Wu.
\newblock Efficient retrieval of the top-k most relevant spatial web objects.
\newblock {\em Proceedings of the VLDB Endowment}, 2(1):337--348, 2009.

\bibitem{cui2013online}
W.~Cui, Y.~Xiao, H.~Wang, Y.~Lu, and W.~Wang.
\newblock Online search of overlapping communities.
\newblock In {\em Proceedings of the 2013 ACM SIGMOD international conference
  on Management of data}, pages 277--288, 2013.

\bibitem{cui2014local}
W.~Cui, Y.~Xiao, H.~Wang, and W.~Wang.
\newblock Local search of communities in large graphs.
\newblock In {\em Proceedings of the 2014 ACM SIGMOD international conference
  on Management of data}, pages 991--1002. ACM, 2014.

\bibitem{de2008keyword}
I.~De~Felipe, V.~Hristidis, and N.~Rishe.
\newblock Keyword search on spatial databases.
\newblock In {\em 2008 IEEE 24th International Conference on Data Engineering},
  pages 656--665. IEEE, 2008.

\bibitem{expert2011uncovering}
P.~Expert, T.~S. Evans, V.~D. Blondel, and R.~Lambiotte.
\newblock Uncovering space-independent communities in spatial networks.
\newblock {\em Proceedings of the National Academy of Sciences},
  108(19):7663--7668, 2011.

\bibitem{fang2017effective}
Y.~Fang, R.~Cheng, X.~Li, S.~Luo, and J.~Hu.
\newblock Effective community search over large spatial graphs.
\newblock {\em Proceedings of the VLDB Endowment}, 10(6):709--720, 2017.

\bibitem{fang2016effective}
Y.~Fang, R.~Cheng, S.~Luo, and J.~Hu.
\newblock Effective community search for large attributed graphs.
\newblock {\em Proceedings of the VLDB Endowment}, 9(12):1233--1244, 2016.

\bibitem{fang2018spatial}
Y.~Fang, Z.~Wang, R.~Cheng, X.~Li, S.~Luo, J.~Hu, and X.~Chen.
\newblock On spatial-aware community search.
\newblock {\em IEEE Transactions on Knowledge and Data Engineering},
  31(4):783--798, 2018.

\bibitem{fang2020effective}
Y.~Fang, Y.~Yang, W.~Zhang, X.~Lin, and X.~Cao.
\newblock Effective and efficient community search over large heterogeneous
  information networks.
\newblock {\em Proceedings of the VLDB Endowment}, 13(6):854--867, 2020.

\bibitem{fortunato2010community}
S.~Fortunato.
\newblock Community detection in graphs.
\newblock {\em Physics reports}, 486(3-5):75--174, 2010.

\bibitem{girvan2002community}
M.~Girvan and M.~E. Newman.
\newblock Community structure in social and biological networks.
\newblock {\em Proceedings of the national academy of sciences},
  99(12):7821--7826, 2002.

\bibitem{guo2008regionalization}
D.~Guo.
\newblock Regionalization with dynamically constrained agglomerative clustering
  and partitioning (redcap).
\newblock {\em International Journal of Geographical Information Science},
  22(7):801--823, 2008.

\bibitem{hariharan2007processing}
R.~Hariharan, B.~Hore, C.~Li, and S.~Mehrotra.
\newblock Processing spatial-keyword (sk) queries in geographic information
  retrieval (gir) systems.
\newblock In {\em 19th International Conference on Scientific and Statistical
  Database Management (SSDBM 2007)}, pages 16--16. IEEE, 2007.

\bibitem{hristidis2003efficient}
V.~Hristidis, Y.~Papakonstantinou, and L.~Gravano.
\newblock Efficient ir-style keyword search over relational databases.
\newblock In {\em Proceedings 2003 VLDB Conference}, pages 850--861. Elsevier,
  2003.

\bibitem{huang2017attribute}
X.~Huang and L.~V. Lakshmanan.
\newblock Attribute-driven community search.
\newblock {\em Proceedings of the VLDB Endowment}, 10(9):949--960, 2017.

\bibitem{huang2015approximate}
X.~Huang, L.~V. Lakshmanan, J.~X. Yu, and H.~Cheng.
\newblock Approximate closest community search in networks.
\newblock {\em arXiv preprint arXiv:1505.05956}, 2015.

\bibitem{islam2019keyword}
M.~Islam, M.~E. Ali, Y.-B. Kang, T.~Sellis, F.~M. Choudhury, et~al.
\newblock Keyword aware influential community search in large attributed
  graphs.
\newblock {\em arXiv preprint arXiv:1912.02114}, 2019.

\bibitem{kacholia2005bidirectional}
V.~Kacholia, S.~Pandit, S.~Chakrabarti, S.~Sudarshan, R.~Desai, and
  H.~Karambelkar.
\newblock Bidirectional expansion for keyword search on graph databases.
\newblock In {\em Proceedings of the 31st international conference on Very
  large data bases}, pages 505--516, 2005.

\bibitem{lancichinetti2011limits}
A.~Lancichinetti and S.~Fortunato.
\newblock Limits of modularity maximization in community detection.
\newblock {\em Physical review E}, 84(6):066122, 2011.

\bibitem{li2005trip}
F.~Li, D.~Cheng, M.~Hadjieleftheriou, G.~Kollios, and S.-H. Teng.
\newblock On trip planning queries in spatial databases.
\newblock In {\em International symposium on spatial and temporal databases},
  pages 273--290. Springer, 2005.

\bibitem{li2014efficient}
G.~Li, S.~Chen, J.~Feng, K.-l. Tan, and W.-s. Li.
\newblock Efficient location-aware influence maximization.
\newblock In {\em Proceedings of the 2014 ACM SIGMOD international conference
  on Management of data}, pages 87--98, 2014.

\bibitem{li2008ease}
G.~Li, B.~C. Ooi, J.~Feng, J.~Wang, and L.~Zhou.
\newblock Ease: an effective 3-in-1 keyword search method for unstructured,
  semi-structured and structured data.
\newblock In {\em Proceedings of the 2008 ACM SIGMOD international conference
  on Management of data}, pages 903--914, 2008.

\bibitem{li2017most}
J.~Li, X.~Wang, K.~Deng, X.~Yang, T.~Sellis, and J.~X. Yu.
\newblock Most influential community search over large social networks.
\newblock In {\em 2017 IEEE 33rd International Conference on Data Engineering
  (ICDE)}, pages 871--882. IEEE, 2017.

\bibitem{li2015influential}
R.-H. Li, L.~Qin, J.~X. Yu, and R.~Mao.
\newblock Influential community search in large networks.
\newblock {\em Proceedings of the VLDB Endowment}, 8(5):509--520, 2015.

\bibitem{Li12}
Y.~Li, D.~Wu, J.~Xu, B.~Choi, and W.~Su.
\newblock Spatial-aware interest group queries in location-based social
  networks.
\newblock In {\em Proc. of the ACM International Conference on Information and
  Knowledge Management}, 2012.

\bibitem{liu2009topic}
Y.~Liu, A.~Niculescu-Mizil, and W.~Gryc.
\newblock Topic-link lda: joint models of topic and author community.
\newblock In {\em proceedings of the 26th annual international conference on
  machine learning}, pages 665--672, 2009.

\bibitem{nallapati2008joint}
R.~M. Nallapati, A.~Ahmed, E.~P. Xing, and W.~W. Cohen.
\newblock Joint latent topic models for text and citations.
\newblock In {\em Proceedings of the 14th ACM SIGKDD international conference
  on Knowledge discovery and data mining}, pages 542--550, 2008.

\bibitem{newman2004finding}
M.~E. Newman and M.~Girvan.
\newblock Finding and evaluating community structure in networks.
\newblock {\em Physical review E}, 69(2):026113, 2004.

\bibitem{qin2009querying}
L.~Qin, J.~X. Yu, L.~Chang, and Y.~Tao.
\newblock Querying communities in relational databases.
\newblock In {\em 2009 IEEE 25th International Conference on Data Engineering},
  pages 724--735. IEEE, 2009.

\bibitem{sozio2010community}
M.~Sozio and A.~Gionis.
\newblock The community-search problem and how to plan a successful cocktail
  party.
\newblock In {\em Proceedings of the 16th ACM SIGKDD international conference
  on Knowledge discovery and data mining}, pages 939--948. ACM, 2010.

\bibitem{wang2012truss}
J.~Wang and J.~Cheng.
\newblock Truss decomposition in massive networks.
\newblock {\em Proceedings of the VLDB Endowment}, 5(9):812--823, 2012.

\bibitem{xu2012model}
Z.~Xu, Y.~Ke, Y.~Wang, H.~Cheng, and J.~Cheng.
\newblock A model-based approach to attributed graph clustering.
\newblock In {\em Proceedings of the 2012 ACM SIGMOD international conference
  on management of data}, pages 505--516, 2012.

\bibitem{yang2012socio}
D.-N. Yang, C.-Y. Shen, W.-C. Lee, and M.-S. Chen.
\newblock On socio-spatial group query for location-based social networks.
\newblock In {\em Proceedings of the 18th ACM SIGKDD international conference
  on Knowledge discovery and data mining}, pages 949--957, 2012.

\bibitem{yuan2016rsknn}
Y.~Yuan, X.~Lian, L.~Chen, Y.~Sun, and G.~Wang.
\newblock Rsknn: knn search on road networks by incorporating social influence.
\newblock {\em IEEE Transactions on Knowledge and Data Engineering},
  28(6):1575--1588, 2016.

\bibitem{Yuan16}
Y.~Yuan, X.~Lian, L.~Chen, Y.~Sun, and G.~Wang.
\newblock Rs$k$nn: $k$nn search on road networks by incorporating social
  influence.
\newblock {\em {IEEE} Trans. Knowl. Data Eng.}, 28(6), 2016.

\bibitem{zhang2013combining}
W.~Zhang, J.~Wang, and W.~Feng.
\newblock Combining latent factor model with location features for event-based
  group recommendation.
\newblock In {\em Proceedings of the 19th ACM SIGKDD international conference
  on Knowledge discovery and data mining}, pages 910--918, 2013.

\bibitem{zhang2019keyword}
Z.~Zhang, X.~Huang, J.~Xu, B.~Choi, and Z.~Shang.
\newblock Keyword-centric community search.
\newblock In {\em 2019 IEEE 35th International Conference on Data Engineering
  (ICDE)}, pages 422--433. IEEE, 2019.

\bibitem{zhou2009graph}
Y.~Zhou, H.~Cheng, and J.~X. Yu.
\newblock Graph clustering based on structural/attribute similarities.
\newblock {\em Proceedings of the VLDB Endowment}, 2(1):718--729, 2009.

\end{thebibliography}

\end{document}